\documentclass[twoside]{article}
\usepackage{xcolor}
\usepackage{graphicx}
\usepackage{hyperref}
\usepackage{amsthm}
\usepackage{amssymb}
\usepackage{algorithm}
\usepackage{algpseudocode}
\usepackage{tcolorbox}
\usepackage{graphicx}
\usepackage[utf8]{inputenc}
\usepackage{multirow} 
\usepackage{amsmath}
\usepackage{booktabs}
\usepackage{wrapfig} 

\usepackage[accepted]{aistats2025}
\usepackage{natbib}

\newtheorem{theorem}{Theorem}[section]

\newtheorem{lemma}[theorem]{Lemma}
\newtheorem{proposition}[theorem]{Proposition}
\newtheorem{corollary}[theorem]{Corollary}

\DeclareMathOperator*{\argmin}{arg\,min}

\begin{document}

%
\runningtitle{Your copula is a classifier in disguise}

%

\twocolumn[

\aistatstitle{Your copula is a classifier in disguise:\\
classification-based copula density estimation}

\aistatsauthor{ David Huk \And Mark Steel \And Ritabrata Dutta}

\aistatsaddress{ University of Warwick } ]

\begin{abstract}
  We propose reinterpreting copula density estimation as a discriminative task. Under this novel estimation scheme, we train a classifier to distinguish samples from the joint density from those of the product of independent marginals, recovering the copula density in the process. We derive equivalences between well-known copula classes and classification problems naturally arising in our interpretation. Furthermore, we show our estimator achieves theoretical guarantees akin to maximum likelihood estimation. By identifying a connection with density ratio estimation, we benefit from the rich literature and models available for such problems. Empirically, we demonstrate the applicability of our approach by estimating copulas of real and high-dimensional datasets, outperforming competing copula estimators in density evaluation as well as sampling.   \end{abstract}

\section{INTRODUCTION}

\begin{figure*}[t]
    \centering
    \includegraphics[width=0.8\textwidth]{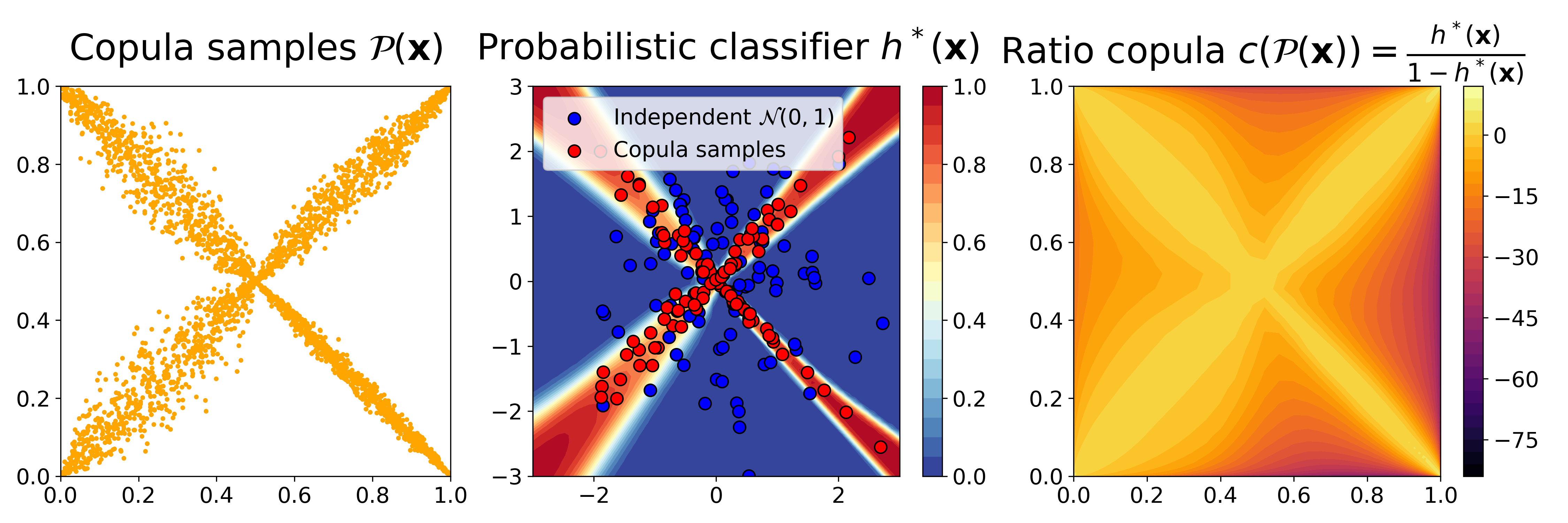}
    \caption{Our ratio copula model estimates copula densities through classification. First panel: dependent copula data. Second panel: we transform the dependent copula data to a Gaussian scale and train a classifier to distinguish between dependent data and an independent Gaussian noise. Third panel: this classifier learns to identify dependence and ultimately recovers the copula density, shown on the log scale above.}
    \label{fig:enter-label}
\end{figure*}
Copulas are a popular tool for modelling multivariate densities by decomposing the estimation procedure into two steps: first, learning the marginal density of each variable; and second, stitching those marginals together through a copula model (the word copula comes from Latin and means “a link, tie, bond”). The adoption of copulas is primarily due to their distinct advantages over other models as a consequence of this two-step procedure. Namely, marginal models can be purpose-built for each dimension, which is relevant for applications, {\it e.g.}~in neuroscience \citep{berkes2008characterizing}, climate research \citep{huk2023probabilistic}, econometrics \citep{czado2022vinefin}, and domain adaptation \citep{lopez2012semi}. Further, copulas focus all optimisation efforts uniquely on modelling dependence, leading to higher quality densities and samples \citep{tagasovska2019copulas}, and allows for the estimation of marginal models in parallel or in real time as used in Bayesian density estimation \citep{huk2024quasi}, Bayesian Networks \citep{elidan2010copula}, and reinforcement learning \citep{wang2021multi}. 

However, current copula models offer a limited selection of efficient methods depending on the use-case, with Gaussian and vine copulas being the two main candidates. The former, whilst fast, is often too simplistic to correctly capture dependence structures. The latter has an exponentially large model space scaling with the dimension, leading to greedy model searches used in practice, and further requires a simplifying assumption on the dependence structure itself to be computationally feasible. Unequivocally, the aforementioned application domains would benefit from enhanced copula models, able to scale to high dimensions, and simultaneously capable of capturing complex patterns while being computationally achievable.

To this aim, we leverage a novel connection between copula densities and classification to propose a new copula estimation framework. This involves training a classifier to distinguish observed dependent data from an independent version of itself, recovering the true copula density at optimality. After providing a background on relevant topics in Section~\ref{sec:background}, we elicit the connection between copulas and classification in Section~\ref{sec:Ratiocop} and develop our classification-based copula estimation scheme. We show equivalences between copulas classifications in Section~\ref{sec:equivalences}. We empirically validate our approach in Section~\ref{sec:Experiments}, and discuss related works in Section~\ref{sec:rel_work} with our conclusion in Section~\ref{sec:diss}.

\section{BACKGROUND}
\label{sec:background}

\paragraph{Notation.} Let $p(\mathbf{x})$ be a multivariate density over $\mathcal{X}\subseteq\mathbb{R}^d$, from which we observe i.i.d.~samples $\mathcal{D}_P=\{\mathbf{x}^{k}\}_{k=1}^{T} \sim p(\mathbf{x})$. Similarly, let $p_1(x_1),\ldots,p_d(x_d)$ be the marginal densities of $p(\mathbf{x})$, over $\mathcal{X}_i\subseteq \mathbb{R}$ with corresponding i.i.d.~samples $\mathcal{D}_{P_i}=\{{x_i}^{k}\}_{k=1}^{T} \sim p_i({x_i})$ for $i=1,\ldots,d$, and assume that all distributions are continuous. Further, let $P$ and $P_1,\ldots,P_d$ be the respective cumulative distribution functions (CDFs) of the previously mentioned densities, in order of appearance. We define applications of CDFs such as $\mathcal{P}(\mathbf{x})\equiv (P_1(x_1),\ldots,P_d(x_d))$ as shorthand for the dimension-wise application of marginal CDFs to a vector or $\mathcal{P}(\mathcal{D})$ to mean their application to every vector element of a set. Probability and cardinality of a set $A$ are denoted by $\mathbb{P}(A)$ and $|A|$ correspondingly. 

\subsection{Copula-Based Density Estimation}
Sklar's theorem (\cite{sklar} found in Appendix \ref{apdx:copulas}) states that any joint probability density function with continuous marginal densities can always be uniquely factorised into the product of its marginals and a copula density $c:[0,1]^d\mapsto(0,\infty)$ in the following way:
\begin{eqnarray}
\label{eq:copula}
    p(\mathbf{x})
    &=& p(x_{1}, \ldots, x_{d})\\\nonumber
    &=& \prod_{i=1}^d p_i(x_i)\cdot {c}(P_{1}(x_{1}), \ldots, P_{d}(x_{d}))\\\nonumber
     &=& \prod_{i=1}^d p_i(x_i)\cdot {c}(\mathcal{P}(\mathbf{x}))
\end{eqnarray}
Consequently, one can decompose the estimation problem of learning $p(\mathbf{x})$ into first learning all marginals $\{p_i\}_{i=1}^d$, and then inferring an appropriate copula model for the dependence structure of $\mathcal{P}(\mathbf{x}) \in [0,1]^d$.

For example, a scientist wishes to model daily temperature data over thousands of stations in different parts of a continent. With their domain expertise, they can design an effective model $p_i(x_i|S_i)$ for the temperature $x_i$ of a given station $i$, taking into account relevant information $S_i$ such as the past days' observed temperature and other variables such as rainfall, cloud cover and humidity in the area. Such a marginal model $p_i$ can be constructed for each station separately as it becomes a unidimensional time-series model with predictors. However, modelling all locations jointly while conditioning on every predictor across time and space might be unmanageable. In such cases, a copula is the appropriate tool to stitch the separate station-specific models together into a mathematically valid joint model over the whole space using \eqref{eq:copula}. 

Whereas taking the product of marginal models $\prod p_i(x_i)$ would also give a valid density model, the joint behaviour would be unaccounted for (represented as an independence copula giving value 1 to all possible observations $\mathcal{P}(\mathbf{x})$), effectively assuming independence across all stations. This is an issue since, for instance, the temperature at two nearby stations should be similar - more than that of two stations located further apart. This is exactly the joint behaviour that a copula serves to capture. 

\paragraph{Current copula models:} There exist many types of copula models, both parametric and non-parametric. Most parametric copulas are only really suited for modelling 2D dependences and greatly suffer from the curse of dimensionality \citep{hofert2013archimedean}. The Gaussian copula (see Appendix \ref{apdx:copulas_gauss}) is a popular parametric choice as it can be fitted quickly even to moderate dimensions. However, it lacks the desired flexibility to capture complex patterns or heavy tails; one of the argued reasons for the 2008 financial crash \citep{salmon2012formula}. Archimedian and Archimax parametric copula families \citep{ng2022inference,gorecki2024hierarchical} have similar drawbacks. Among non-parametric copulas, kernel density estimator (KDE) copulas \citep{geenens2017probit} are commonly used. They transform the observed data via appropriate (often Gaussian) inverse CDFs to a latent space and perform regular KDE on the latent density. However, doing so suffers from drawbacks common to KDE methods in higher dimensions. Deep learning copula models are still a nascent line of research, with a handful of candidate solutions such as \cite{kamthe2021copula,janke2021implicit,ashok2024tactis}. The Implicit Generative Copula (IGC) of \cite{janke2021implicit} is a promising model, which works by generating data in a latent space and applying an approximate integral transformation to recover samples in $[0,1]$. As an implicit model only able to generate samples, the IGC cannot evaluate densities, thus preventing many applications of copula models. As such,  current copula models supporting density estimation and sampling are mostly limited to low or medium-dimensional modelling. 

\paragraph{Vine copulas:} The state-of-the-art approach is vine copula models, offering a divide-and-conquer solution to expand copula models to higher dimensions. Vine copulas build a hierarchical decomposition of the joint copula density by conditional factorisation into bivariate sub-components along dimensions (see Appendix \ref{apdx:vine} and \cite{aas2009pair,czado2019analyzing,czado2022vine} for more detail). In practice, it is common to ignore part of the conditional relationships in an approach termed simplified vines \citep{nagler2016evading, nagler2017nonparametric}. This results in a convergence rate that is independent of the dimension so long as the true data-generating model is a simplified vine, and its correct decomposition has been identified \citep{nagler2016evading}. However, vines have an exponentially large model space with the dimension, meaning greedy search algorithms are employed in practice. Already for 5-dimensional data, itself simulated from a vine copula, recovering the correct decomposition with current vine models is extremely challenging \citep{janke2021implicit}, meaning these models are often misspecified in practice.

\subsection{Probabilistic Classification for Density Ratio Estimation}
\label{sec:DRE_class}

The problem of density ratio estimation (DRE) is a crucial inference problem in machine learning and statistics \citep{sugiyama2010density}, most notably framed as a probabilistic classification task between samples from both densities involved in the ratio \citep{gutmann2010noise,gutmann2011bregman,sugiyama2012density,menon2016linking}.

Given a dataset $\mathcal{D}_{p}=\{\mathbf{x}^{k}_{p}\}_{k=1}^{T_p} \sim p(\mathbf{x})$ and $\mathcal{D}_{q}=\{\mathbf{x}^{k}_{q}\}_{k=1}^{T_q} \sim q(\mathbf{x})$ with $q(\mathbf{x})>0$ whenever $p(\mathbf{x})>0$, we are interested in estimating the ratio $r(\mathbf{x})= p(\mathbf{x})/q(\mathbf{x})$. Fictitious labels $y$ are assigned so that $y=1$ for $\mathcal{D}_{p}$ and $y=0$ for $\mathcal{D}_{q}$, with identical prior odds being assumed (although this can be relaxed). Then a probabilistic classifier $h(\mathbf{x}): \mathcal{X}\rightarrow [0,1]$ is trained to assign the probability of $\mathbf{x}$ being simulated from $p(\mathbf{x})$. By Bayes' rule, the optimal value of $h(\mathbf{x})$ is  $h^*(\mathbf{x})=\mathbb{P}(y=1|\mathbf{x})$, known as a {\it Bayes optimal classifier}  \citep{hastie2009elements,sugiyama2012density,gutmann2012noise}. We denote $\frac{h(\mathbf{x})}{1-h(\mathbf{x})}$ as the {\it decision boundary} since if this quantity is above one, it indicates $\mathbf{x}$ is more likely to belong to class $y=1$ than $y=0$. Crucially, the Bayes classifier can be equated to a density ratio via the decision boundary as follows:
\begin{eqnarray}
\label{eq:bayesoptim_ratio}
r(\mathbf{x})=\frac{p(\mathbf{x})}{q(\mathbf{x})}=\frac{p(\mathbf{x} \mid y=1)}{q(\mathbf{x} \mid y=0)}=\frac{h^*(\mathbf{x})}{1-h^*(\mathbf{x})}.
\end{eqnarray}
Developing methods for such problems is an active area of research, with new models being proposed for increasingly complex tasks \citep{rhodes2020telescoping,choi2021featurized,choi2022density}.

\section{PROBABILISTIC CLASSIFICATION FOR COPULA ESTIMATION}
\label{sec:Ratiocop}

In what follows, after establishing the connection between copula estimation and probabilistic classification, we derive a novel {\it ratio copula} estimator. Throughout, we assume that marginal densities are known and kept fixed, thus for $\mathbf{x}\sim p(\mathbf{x})$, we treat $\mathcal{P}(\mathbf{x})$  as samples from the true copula. 
This is a common approach in copula estimation \citep{fermanian2005goodness,geneve2003research,scaillet2007kernel}, and ensures the validity of the resulting copula-based density (\cite{ashok2024tactis} Proposition 2) even in non-parametric settings.  

We begin by noticing that, using Sklar's theorem (\ref{eq:copula}), a copula density can be re-written as a
ratio between the joint density and the product of the marginals,
\begin{eqnarray}
\label{eq:copula_frac}
c(\mathcal{P}(\mathbf{x})) =
\frac{p(x_1,\ldots,x_d)}{\prod_i p_i(x_i)}. 
\end{eqnarray}

Then, following the identity of Equation~\ref{eq:bayesoptim_ratio}, we can express our copula density $c(\mathbf{\mathcal{P}(\mathbf{x})}):[0,1]^d\rightarrow\mathbb{R}^+$ as, \begin{eqnarray}
\label{eq:ratio_cop}
c(\mathcal{P}(\mathbf{x})) = \frac{h^*(\mathbf{x})}{1-h^*(\mathbf{x})}, 
\end{eqnarray}  
where $h^*(x)$ is the optimal Bayes classifier for $\mathcal{D}_p \sim p(\mathbf{x})$ and $\mathcal{D}_{\prod_i p_i} \sim \prod_{i=1}^dp_i(x_i)$, assuming unitary prior odds. Under this interpretation, a copula function acts like a probabilistic classifier on $\mathcal{X}$, identifying regions of $\mathcal{X}$ where samples are more likely to belong to the joint distribution than the product of marginals. In other words, copulas are classifiers trained to identify dependence. 

\subsection{Ratio Copula Estimator}
\label{sec:method}

We propose to train a {\it ratio copula} estimator $c^{\text{ratio}} $ for the true copula density function $c$ using probabilistic classification. However, estimating the Bayes optimal classifier on $\mathcal{X}$ might not be straightforward depending on the constraints of the space, and estimating it on $[0,1]^d$ directly is even more cumbersome due to the constrained support. For this reason, we use the probit  transformation $\Phi^{-1}:[0,1]\mapsto\mathbb{R}$ applied dimension-wise, commonly used for copula models \citep{geenens2017probit}):
\begin{equation}
\label{eq:transform}
  \Phi^{-1}\circ \mathcal{P}(\mathbf{x}): \mathcal{X}\rightarrow \mathbb{R}^d.
\end{equation}
By the probability integral transformation\footnote{ If $x_i\sim p_i$ is continuous, then $P_i(x_i)\sim Unif[0,1].$
}, if $\mathbf{x}\sim p(\mathbf{x})$, then $\Phi^{-1}\circ \mathcal{P}(\mathbf{x})$ marginally follows a standard Gaussian. Therefore, the product of their marginal densities is $\tilde q(\cdot)= \prod_{i=1}^d\mathcal{N}(\cdot;0,1)$. Nevertheless, their joint density will be some other unknown function $\tilde p (\Phi^{-1}\circ \mathcal{P}(\mathbf{x}))$. By Sklar's theorem, $\tilde p$ admits a copula decomposition with a corresponding copula $\tilde c$. Since the marginals of $\tilde p$ follow a standard Gaussian, its marginal CDFs are exactly $\Phi$. Consequently, the copula expression simplifies, having the marginal CDFs $\Phi$ of $\tilde p$ cancelling with the $\Phi^{-1}$ part of our transformation in (\ref{eq:transform}):
\begin{equation}
    \tilde c(\Phi\left(\Phi^{-1}\circ \mathcal{P}(\mathbf{x})\right))=\tilde c(\mathcal{P}(\mathbf{x})).
\end{equation}
By the uniqueness of copula densities, as $\tilde c$ and $c$ describe the same random variables, we recover our original copula $c(\mathcal{P}(\mathbf{x}))$ with an expression in terms of the transformed densities:
\begin{equation}
\label{eq:cop_on_gauss}
    c(\mathcal{P}(\mathbf{x})) = \frac{\tilde p(\Phi^{-1}\circ \mathcal{P}(\mathbf{x}))}{\tilde q (\Phi^{-1}\circ \mathcal{P}(\mathbf{x}))} = \frac{\tilde p ( \Phi^{-1}\circ \mathcal{P}(\mathbf{x}))}{\prod_{i=1}^d\mathcal{N}( \Phi^{-1}\circ P_i(x_i);0,1)}.
\end{equation}
Thus, we can now focus on the ratio of the RHS in (\ref{eq:cop_on_gauss}). Namely, our aim becomes training a classifier to distinguish dependent data $\mathcal{D}_{\tilde p}=\Phi^{-1}\circ\mathcal{P}(\mathcal{D}_p)$ from independent data $\mathcal{D}_{\tilde q}=\Phi^{-1}\circ\mathcal{P}(\mathcal{D}_{\prod_i p_i})$, with respective labels $y=1$ and $y=0$, all conveniently on $\mathbb{R}^d$.

Following the literature on probabilistic classification, we assume $c$ belongs to a parametric class of functions $\mathcal{C}\equiv \{c^{\text{ratio}}(\cdot;\boldsymbol{\theta}) | c^{\text{ratio}}( \cdot; \boldsymbol{\theta}):\mathbb{R}^d\rightarrow\mathbb{R}^+, \boldsymbol{\theta}\in \Theta \}$, which can consist of simple functions ({\it e.g.}~linear) to more complex functions ({\it e.g.}~non-linear functions parametrized by suitable neural network architectures).
Then, we estimate the parameters of $c^{\text{ratio}}(\Phi^{-1}\circ\mathcal{P}(\mathbf{x});\boldsymbol{\theta})$ by minimizing the logistic loss:
\begin{eqnarray}
\label{eq:loss}
\widehat{\boldsymbol{\theta}} &=& \argmin_{\boldsymbol{\theta}\in\Theta}\mathcal{L}(\boldsymbol{\theta})\\ \nonumber
 &=& \argmin_{\boldsymbol{\theta}\in\Theta}
    -\mathbb{E}_{\tilde p} \ln \left(\frac{{c^{\text{ratio}}}(\Phi^{-1}\circ\mathcal{P}(\mathbf{x});\boldsymbol{\theta})}{\nu+{c^{\text{ratio}}}(\Phi^{-1}\circ\mathcal{P}(\mathbf{x});\boldsymbol{\theta})}\right)\\\nonumber
    && \quad \quad\,-\nu \cdot\mathbb{E}_{\tilde q} \ln \left(\frac{\nu}{\nu+{c^{\text{ratio}}}(\Phi^{-1}\circ\mathcal{P}(\mathbf{x}) ; \boldsymbol{\theta})}\right)
\end{eqnarray}
where $\nu= \mathbb{P}(y=0)/\mathbb{P}(y=1)$ to compensate for unequal prior odds. The ratio copula $c^{\text{ratio}}(\cdot;\boldsymbol{\theta})$ implicitly defines a classifier between $\mathcal{D}_{\tilde p}$ and $\mathcal{D}_{\tilde q}$. As minimising $\mathcal{L}(\boldsymbol{\theta})$ is equivalent to minimising a Bregman divergence between $\mathcal{C}$ and 
${c}(\cdot)$ \citep{menon2016linking}, 
$\exists \ \theta^*$ such that
${c^{\text{ratio}}}(\cdot;\boldsymbol{\theta}^*)$ is the projection of $c(\cdot)$ on our model class $\mathcal{C}$. If our model class is well specified,  then
$c^{\text{ratio}}(\cdot;\theta^*) = c(\cdot)$. In such a case, our estimator will asymptotically recover ${c^{\text{ratio}}}(\cdot;\boldsymbol{\theta}^*)$, without needing any normalising constant \citep{gutmann2012noise}. However, as this only holds at optimality, we simultaneously estimate a normalising constant as an additional parameter $Z$ included in $\boldsymbol{\theta}$, which pre-multiplies the ratio copula. The estimation of $Z$ is done jointly with $\boldsymbol{\theta}$ by minimising the same loss, ensuring that the resulting ratio copula is a valid density.

Our ratio copula estimator inherits the theoretical guarantees associated with the loss in (\ref{eq:loss}) from \cite{gutmann2012noise}, which we concretise here with a justification of its applicability given in Appendix \ref{apdx:thm}.
\begin{theorem} 
\label{thm:cop}
\textbf{Estimator properties; \cite{gutmann2012noise}} Assuming continuity of all marginal densities, ${c^{\text{ratio}}}(\Phi^{-1}\circ\mathcal{P}(\mathbf{x});\boldsymbol{\theta}^*)$ obtained by minimising Equation (\ref{eq:loss}) is unique. Further, under certain regularity conditions detailed in Appendix \ref{apdx:thm}, the copula estimator is consistent with $\boldsymbol{\hat{\theta}}\overset{\mathbb{P}}{\to}\boldsymbol{\theta}^*$, and $\sqrt{|\mathcal{D}_p|}(\boldsymbol{\hat{\theta}}-\boldsymbol{\theta}^*)$ is asymptotically normal.
\end{theorem}

Following the analysis from \cite{gutmann2012noise} (Corollaries 4-7) of the loss function (\ref{eq:loss}), our copula estimator's accuracy is influenced by two main parts: the factor of independent to dependent observations $\nu$ and the form of $\tilde q$. Specifically, we want as many samples from $\tilde q$ as possible in order to inflate $\nu$, and we want $\tilde q$ to be similar to $\tilde{p}$. However, in our case, only dependent observations $\mathcal{D}_{\tilde p}$ are initially available and the form of $\tilde q$ is necessarily fixed by virtue of estimating a copula. 

\subsection{Learning ratio copulas in practice}
In the following, we demonstrate how to leverage our construction of (\ref{eq:cop_on_gauss}) and our formulation in terms of density ratio estimation to satisfy our desiderata on the sampling and density of $\tilde q$.
\paragraph{Efficient sampling of ${\tilde q}$:}During training, all observations are labelled as dependent with $y=1$, whereas we have no independent samples from ${\tilde q}$. To obtain such samples, one could
use the heuristic of recreating independent samples
by dimension-wise shuffling the values of $\mathcal{D}_{\tilde p}$, but this
would lead to a misrepresentation of the statistical
properties. Instead, we can take advantage of our transformation to $\mathbb{R}^d$ in (\ref{eq:transform}) and exploit the fact that ${\tilde q} (\Phi^{-1}\circ\mathcal{P}(\mathbf{x}))= \mathcal{N}(\Phi^{-1}\circ\mathcal{P}(\mathbf{x});\mathbf{0},\mathbf{I}_d)$, which we can sample from {\it ad nauseam}. This satisfies our first desideratum, namely, access to a large amount of samples from $\tilde q$.

\paragraph{Narrowing the KL gap in a copula:} For the second desideratum, we want the density of the independent data to be close to that of the dependent data. We can rewrite the density ratio in ($\ref{eq:cop_on_gauss}$) by introducing an intermediate density:
\begin{eqnarray}
\label{eq:gg_ratio}
&&c(\mathcal{P}(\mathbf{x})) = \frac{\tilde p ( \Phi^{-1}\circ \mathcal{P}(\mathbf{x}))}{\mathcal{N}_d( \Phi^{-1}\circ \mathcal{P}(\mathbf{x});\mathbf{0},\mathbf{I}_d)}  \\
    &=&\underbrace{\frac{\tilde{p}\left(\Phi^{-1}\circ \mathcal{P}(\mathbf{x})\right)}
{\mathcal{N}_d(\Phi^{-1}\circ \mathcal{P}(\mathbf{x})|\mathbf{0},\mathbf{\Sigma})}}_{\text{ratio copula model}}
\cdot
\underbrace{\frac{\mathcal{N}_d(\Phi^{-1}\circ \mathcal{P}(\mathbf{x})|\mathbf{0},\mathbf{\Sigma})}
{\mathcal{N}_d(\Phi^{-1}\circ \mathcal{P}(\mathbf{x})|\mathbf{0},\mathbf{I}_d)}}_{\text{known Gaussian ratio}}\nonumber
\end{eqnarray}
which is a centred correlated Gaussian
with covariance matrix $\Sigma$, a good choice being the sample covariance of $\Phi^{-1}\circ \mathcal{P}(\mathcal{D}_p)$ to mimic the inherent dependence in the density $\tilde{p}$. This intermediate Gaussian guides the classifier closer to the data density to improve the accuracy of the final ratio copula. As the analytical expression of the second ratio between the Gaussians is known, we instead, we focus our estimation efforts on the ratio $\frac{\tilde{p}\left(\Phi^{-1}\circ \mathcal{P}(\mathbf{x})\right)}
{\mathcal{N}_d(\Phi^{-1}\circ \mathcal{P}(\mathbf{x})|\mathbf{0},\mathbf{\Sigma})}$  by classifying between the dependent
data $\mathcal{D}_{\tilde p}$ and $\mathbf{z}\sim \mathcal{N}_d(\mathbf{z}|\mathbf{0},\mathbf{\Sigma})$. Once the ratio copula model is trained on this task, we recover the
correct copula by multiplying the trained model with
the ratio of Gaussians. We call the resulting copula estimator a {\it Gaussian-Guided (GG) ratio copula}.

We give an algorithmic description of the proposed approach in Algorithm \ref{alg:ratio_copula} in Appendix \ref{apdx:exp} where orange parts are optional for producing a Gaussian-Guided ratio copula.

Furthermore, it is of interest to measure how difficult a classification task is for a ratio copula model. In particular, knowing when estimating a Gaussian copula might be too hard a task for a non-parametric (or over-parameterised) estimator, would indicate that we should use at least a Gaussian-Guided ratio copula model, to match the gap between densities.

\paragraph{Bound on Kullback-Leibler (KL) divergence for Gaussian copulas.} As a Gaussian copula is the ratio between an independent Gaussian and a Gaussian with a nontrivial covariance matrix $\Sigma$, the difficulty of the classification-based estimation revolves around how different these two densities are. Fortunately, we can recover a simple analytical expression for the KL divergence between the two Gaussians. This is useful to show that as long as the ellipsoid formed by the Gaussian $\Tilde{p}$ is not too narrow (ie. the eigenvalues of its covariance are lower bounded), then the KL is upper bounded by a constant depending on $d$.

\begin{lemma}[\textbf{Bound on KL of a Gaussian copula classifier}]
\label{lemma:KL_gauss}
Consider a $d$-dimensional Gaussian copula with covariance matrix $\Sigma$. Then, the KL divergence between the copula numerator and denominator is upper-bounded with
\begin{eqnarray}
KL\left(\mathcal{N}_d(.;\mathbf{0},\Sigma)||\,\mathcal{N}_d(.;\mathbf{0},\mathbf{I}_d)\right)\leq -\frac{d}{2}\log(e_{min})
\end{eqnarray}
where $e_1,\ldots,e_d$ are the eigenvalues of $\Sigma$ with $e_{min}$ the (possibly joint) smallest eigenvalue of $\Sigma$.
\end{lemma}

We prove this statement in Appendix \ref{apdx:proof_KLgauss}. The result implies that under restrictions on the smallest width of the ellipsoid of $\Tilde{p}$, the KL cannot be infinite for a fixed dimension $d$, and in fact scales linearly. As expected, if $\Sigma$ is close to nonsingular, the KL divergence can still be very large indeed; it is linear in $d$ but $\Sigma$ is likely to become more ill-conditioned (closer to nonsingularity) as $d$ grows. Consequently, we can rely on the empirical findings of \cite{rhodes2020telescoping}, who show classical DRE methods fail for KL differences of over 20 natural logarithms between the two densities. This result provides an informed way of deciding which DRE method to employ for a Gaussian copula based on the sample covariance of $\mathcal{D}_{\tilde p}$; whether to use a simple ratio copula model, empower it with a Gaussian-Guided ratio copula, or employ one of the many advanced DRE frameworks \citep{rhodes2020telescoping,choi2021featurized,choi2022density,srivastava2023estimating}.

\section{EQUIVALENCES BETWEEN CLASSIFIERS AND COPULAS}
\label{sec:equivalences}

In this Section, we reason about copulas in terms of the classification rules they implicitly define in light of Equation \ref{eq:ratio_cop}, providing new insights into their modelling capabilities. Throughout, we work on the Gaussian space with $\Phi^{-1}\circ \mathcal{P}(\mathbf{x}) \in \mathbb{R}^d$ for any $\mathbf{x}\in \mathcal{X}$.\par
Further, in Figure \ref{fig:cop_class} of Appendix \ref{apdx:cop_class}, we make use of the following identity
\begin{equation*}
    c(\mathbf{P(\mathbf{x})})=\frac{h(\mathbf{x})}{1-h(\mathbf{x})} \iff h(\mathbf{x})=\frac{c(\mathbf{P(\mathbf{x})})}{1+c(\mathbf{P(\mathbf{x})})}
\end{equation*}
to illustrate our theory by showing the implied classifiers $h(\mathbf{x})$ for six bivariate parametric copulas.

\subsection{Gaussian Copulas and QDA}
In the case of a Gaussian copula, the density in the numerator is a correlated Gaussian and the denominator is an uncorrelated Gaussian, both centred at $\mathbf{0}$. From Equation \ref{eq:ratio_cop}, it then follows that the classification problem occurs between a multivariate Gaussian $\Tilde{q}$ with identity covariance and a fully dependent Gaussian $\Tilde{p}$; our aim becomes the classification of two Gaussians with different covariance structures. This is exactly the goal of Quadratic Discriminant Analysis (QDA) \citep{hastie2009elements}, which fits a function $h^k(\mathbf{z};\mu_1,\mu_0,\Sigma_1,\Sigma_0):\mathbb{R}^d\mapsto[0,1]$ to classify between $\mathcal{N}(.;\mu_1,\Sigma_1)$ with labels $y=1$ and $\mathcal{N}(.;\mu_0,\Sigma_0)$ with labels $y=0$.

\begin{proposition}[\textbf{Gaussian copulas and QDA}]
\label{prop:cop_qda}
    Consider a Gaussian copula parameterised by a covariance matrix $\Sigma$. The Bayes optimal classifier associated with this copula is found by Quadratic Discriminant Analysis between a standard Gaussian and a centred Gaussian with covariance matrix $\Sigma$.
\end{proposition}
We provide a proof in Appendix \ref{apdx:proofs_cop_qda}. This result opens up a new way of estimating Gaussian copulas by fitting a QDA classifier to $\Phi^{-1}\circ \mathcal{P}(\mathbf{x})$. Further, this result provides a new perspective on Gaussian copulas by linking them to the quadratic decision boundaries of QDA, characterising the complexity of Gaussian copulas. We show an equivalent result for Student's t copulas in Appendix \ref{apdx:proof_student}.

\subsection{In the case of a KDE copula}

Next, we look at KDE copulas regarding their implicit classification problems to establish their relative complexity. In particular, for a Gaussian KDE copula, we can obtain the following: 

\begin{proposition}[\textbf{Gaussian KDE copulas and averages of QDA}]
\label{prop:kde}
    Consider a KDE copula with a Gaussian kernel having identity variance scaled by $b$, based on $T_p$ kernel means at observations $\{(Z_1^k,\ldots,Z_d^k)\}_{k=1}^{T_p}$ for $Z_i^k=\Phi^{-1}\circ P_i(x_i^k)$, of the form:
\begin{align}
&\hat{c}_{T_p}(\Phi^{-1}\circ\mathcal{P}(\mathbf{x}))=  \\
 &\frac{\sum_{k=1}^{T_p} \prod_{i=1}^{d} \mathcal{N}(\Phi^{-1} \circ P_i(x_i) - Z_i^{k} | 0, b)}
{T_p \cdot \mathcal{N}_d\left(\Phi^{-1} \circ \mathcal{P}(\mathbf{x}) | \mathbf{0}, \mathbf{I}_d\right)}.
\end{align}
    Then, the associated Bayes optimal classifier is found by the average of $T_p$ decision boundaries from QDA classifiers $h^k(\mathbf{z};\mu_1^k,\mu_0^k,\Sigma_1,\Sigma_0)$, with component-specific means $\mu_1^k,\mu_0^k$ but shared variances $\Sigma_1,\Sigma_0$ across components.
\end{proposition}

The proof is included in Appendix \ref{apdx:kde}. This can be extended to other types of kernels for similar results. Proposition \ref{prop:kde} describes a KDE copula as an average of QDA decision boundaries over the $T_p$ observed samples. This results in a more complex decision boundary with more nonlinearity introduced linearly in the number of samples. However, due to the curse of dimensionality, an exponential number of samples will be required to maintain a constant level of complexity, thus elucidating KDE copulas' difficulty in scaling to high dimensions.

\subsection{For a simplified vine copula}
Vine copulas are decompositions of copula densities using only bivariate copulas as building blocks and in simplified vine copulas, the bivariate copulas are assumed to have no conditioning (see Appendix \ref{apdx:vine} and \cite{czado2019modelling} for an introduction). Here, we consider two cases, namely, using Gaussian copulas or Gaussian KDE copulas as those building blocks within simplified vine copula models. We give the statement here with a proof found in Appendix \ref{apdx:proof_vine}.

\begin{corollary}
\label{coro:vine}
    The density of a simplified vine copula constructed using bivariate Gaussian copulas as building blocks coincides with the product of $\frac{d(d-1)}{2}$ QDA decision boundaries between two bivariate centred Gaussians with different covariances.\\
    Moreover, the density of a simplified vine copula constructed using Gaussian KDE copulas as building blocks coincides with the product of $\frac{d(d-1)}{2}$ different averages of $T_p$ QDA decision boundaries between two bivariate Gaussians with unequal means and variances.
\end{corollary}
This result is instructive in describing the complexity of simplified vines through their use of multiple decision boundaries. They function much like hierarchical classifiers (examples of which are classification and regression trees, \cite{loh2011classification}, or random forests, \cite{breiman2001random}) by breaking the modelling task into smaller subparts on subgroups of variables. One can also draw a parallel with multilayered perceptrons that have parts of the model, such as their different layers, specialised in capturing specific features of the data relayed to them. Vine copulas instead have many separate models which capture different features, where components further up the decomposition have inputs with more conditioning - encoding the information extracted by previous classifiers. This can then be seen as a mixture of experts model, a common framework in machine learning \citep{yuksel2012twenty,jiang2024mixtral}.

\section{EXPERIMENTS}
\label{sec:Experiments}

\begin{figure*}[t]
    \centering
    \includegraphics[width=0.7\textwidth]{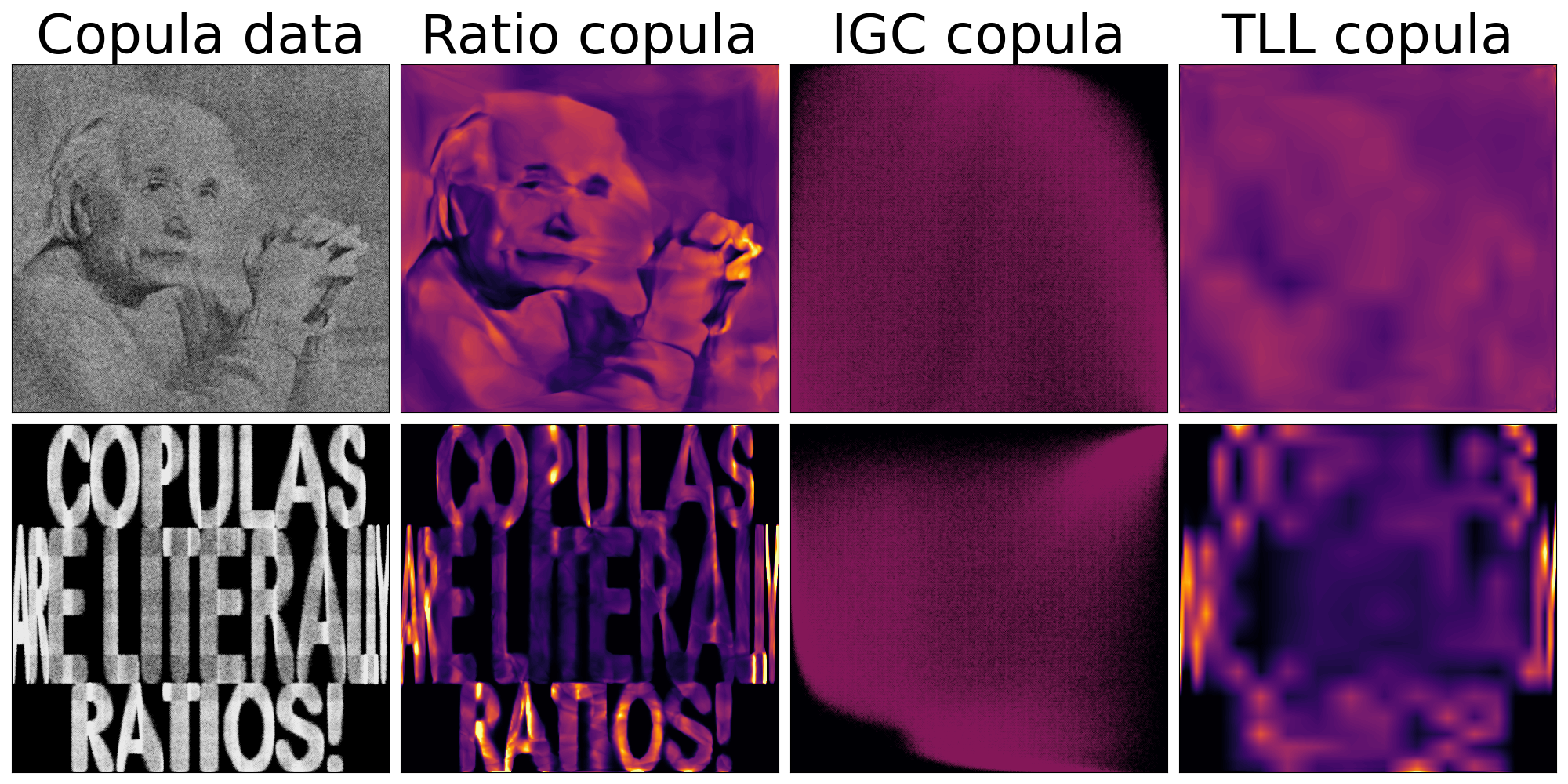}
    \caption{Bivariate nonparametric copula models trained on the dependence in the light of pixels from 2D images. From left to right, columns show on $[0,1]^2$: the observed data, the LL of our ratio copula, samples from the IGC, and the LL of the TLL. Only the ratio copula manages to correctly capture the sharp transitions in the LLs.}
    \label{fig:2d_img}
\end{figure*}

Section \ref{sec:equivalences} described the limitations of popular copulas. To achieve better performance, designing more flexible copula models is essential. In this Section, we experimentally verify the validity of our proposed estimator. Implementation details are discussed in Appendix \ref{apdx:exp} with additional experiments on alternate losses and samples produced by our ratio copula models. Our code is available at \url{https://github.com/Huk-David/Ratio-Copula}. We investigate the following:
\begin{itemize}
    \item Are ratio copulas competitive models in two dimensions, encouraging their use as non-parametric models in vine copula decompositions?
    \item Are ratio copulas effective in high dimensions and on typically challenging data for copula models?
\end{itemize}

\paragraph{Baselines:} We compare our ratio copula to commonly used models for copula estimation. Namely, the Gaussian copula, the TLL\footnote{Transformation local likelihood (TLL), a class of nonparametric copula estimators using (\ref{eq:transform}) with polynomials. } constant, linear or quadratic copulas in the pyvinecopublib package of \cite{vatter20220} as well as a simplified vine copula using TLL constant bivariate copulas. Finally, we compare with the implicit generative copula (IGC) of \cite{janke2021implicit}, as a comparison with an effective deep copula, which is only able to sample.

\paragraph{Metrics:} As a metric for success we choose the {\it average} copula log-likelihood (denoted LL), a well-used metric on its own, but in the context of copula models, once marginal densities have been estimated, only the copula LL is relevant to the LL of the full model. We also compare samples from each copula model using an empirical version of the Wasserstein-2 metric, as done in other probabilistic modelling contributions ({\it e.g.}~\cite{tagasovska2019copulas}).

\paragraph{Sampling:} Our ratio copula offers an analytical density. Hence we use Hamiltonian Monte Carlo \citep{neal2011mcmc}, a gradient-informed MCMC sampler, to output samples from the estimated model.

\begin{figure}[h]
    \centering
    \includegraphics[width=1.\linewidth]{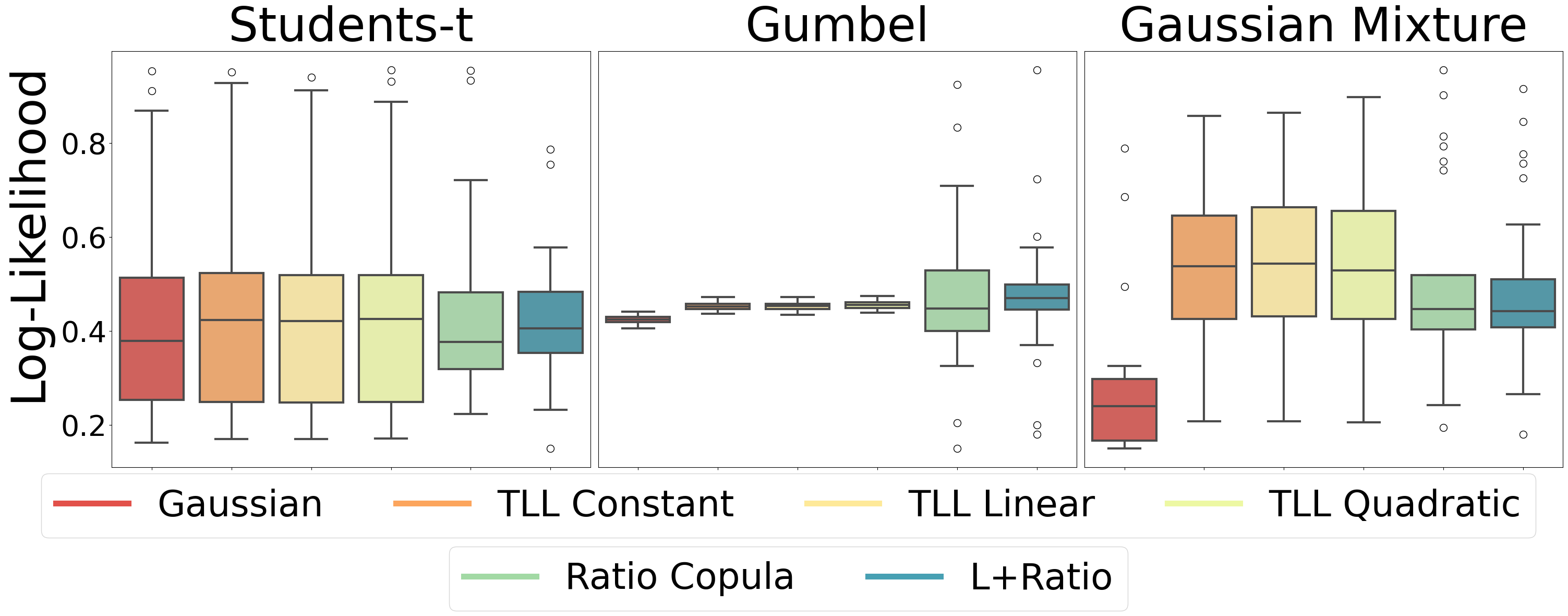}
    \caption{Box plots showing the average LL across 25 fits of bivariate copula models based on samples from the Student's t, Gumbel and Gaussian Mixture parametric copulas.}
    \label{fig:2d_param}
\end{figure}

\paragraph{Estimating 2D copulas:}
Modelling bivariate copulas is particularly valuable, as such models construct vine copula decompositions.

\begin{table*}[t]
\small
\centering
\begin{tabular}{|c|c c|c c|}
\hline
\multirow{2}{*}{Model} & \multicolumn{2}{c|}{Digits} & \multicolumn{2}{c|}{MNIST} \\ \cline{2-5}
                       & LL (Mean $\pm$ Std) & W2 (Mean $\pm$ Std) & LL (Mean $\pm$ Std) & W2 (Mean $\pm$ Std) \\ \hline
GG+ratio               & ${12.97 \pm 0.67}$ & ${8.03 \pm 0.05}$ & ${198.39 \pm 19.72}$ & $36.44 \pm 0.51$ \\ \hline
{GG+Ratio \textit{large}}   & $\mathbf{13.69 \pm 0.77}$ & $\mathbf{7.99 \pm 0.07}$ & $\mathbf{334.42 \pm 45.91}$ & $37.40 \pm 0.79$ \\ \hline
Vine                   & $12.79 \pm 0.07$ & $8.50 \pm 0.02$ & $183.00 \pm 0.39$ & $37.57 \pm 0.18$ \\ \hline
Gaussian               & $11.78 \pm 0.13$ & $8.34 \pm 0.01$ & $126.41 \pm 0.09$ & $37.00 \pm 0.15$ \\ \hline
IGC                    & - & $8.14 \pm 0.03$ & - & $\mathbf{36.02 \pm 0.06}$ \\ \hline
\end{tabular}
\caption{Average LL and W2 scores for the Digits and MNIST datasets with one standard deviation margins over 10 runs, the best scores are in bold. Our method (GG+Ratio) performs best and improves further with a larger network architecture for the classifier (GG+Ratio \textit{large}), showcasing its adaptability.}
\label{Table:highd}
\end{table*}

In Figure \ref{fig:2d_param} we fit parametric copulas based on 5000 samples, namely the  Student-t, the Gumbell and a mixture of two Gaussian copulas with random covariances. We then test the model LL by evaluating 5000 held-out test samples, repeating this experiment 25 times with different seeds. We show the LL values for the Gaussian, TLL copulas, as well as two versions of our ratio copula: Ratio Copula (a multilayered perceptron) and L+Ratio (a polynomial logistic classifier of order 5 with all interaction terms). The Gaussian copula performs the worst overall. The TLL and our models are similar for the Student-t and on the Gumbel but perform slightly worse on the Gaussian mixture. This is a non-trivial achievement since current methods have been extensively optimized for low-dimensional data. Our estimator's variance is higher than the TLL's variance for the Gumbel, likely due to it being an Archimedean copula. Notably, the L+Ratio copula marginally outperforms the Ratio Copula version, which we accredit to its simple structure acting as an implicit regularisation. Both of our models perform consistently well, making them suitable for vine copulas.

In Figure \ref{fig:2d_img}, we show estimates of toy examples of bivariate copulas generated from the dependence found in monochromatic images when data is sampled according to the light of each pixel (more descriptions of this dataset can be found in \ref{apdx:exp_2dimg}). The training size is one million, meant to emulate the big data setting. We show observed data in the first column, with the following columns showing the ratio copula density, samples from the IGC, and the TLL quad density, in that order. The ratio copula captures the sharp transitions of the densities accurately, while the TLL is much more diffuse. The IGC struggles to converge to a correct representation.

\paragraph{High-dimensional examples:}
We model the copula density obtained with a 50/50 train/test split, from the digits dataset ($n=1797, d=64$) and the MNIST dataset ($n=60000,d=784$), both showing handwritten digits. This dataset assesses the ability of our model to pick up complex and multimodal dependencies in the data and is typically difficult for copula models. In Table \ref{Table:highd}, we show the results of the Gaussian-guided ratio copula (GG+Ratio with a simple network and GG+Ratio \textit{large} having a larger network architecture for the classifier, see Appendix \ref{apdx:exp_digits_mnist}), a simplified vine, a Gaussian copula, and the IGC. Unlike the vine copula, whose building blocks scale exclusively with sample size (Corrolary \ref{coro:vine}), our method easily integrates more complex architectures as done with the GG+Ratio \textit{large}. \par
For digits, our method performs best in LL and sampling. For MNIST, we achieve a better LL, and are second in W2 to the IGC. We show generated samples for both digits and MNIST from all methods in Appendix \ref{apdx:exp_digits_mnist}. Notably, samples from the vine on MNIST are of worse quality, and IGC MNIST samples appear over-inflated or unfocused. Only the ratio copula captures the right thickness of the characters.

To conclude, we demonstrated that ratio copulas are extremely flexible models if given enough data (Figure \ref{fig:2d_img}), while in smaller data regimes, the L+ratio might be preferred. Thus, their use is warranted in vine constructions, broadening the class of non-parametric bivariate models. Further, though the application on Digits and MNIST, we demonstrate the ability of ratio copulas to scale to high dimensions with complex dependencies, thereby establishing ratio copulas as a useful non-parametric copula model.

\section{RELATED RESEARCH}
\label{sec:rel_work}
Our research elucidates connections in three ways. Firstly, this paper bridges copulas with DRE and the related Mutual information (MI) estimation. Secondly, in the taxonomy of implicit learning losses from \cite{mohamed2016learning} applied to copulas, our work stands alone, not only through the use of a Bregman divergence, but also as the only deep copula using non-likelihood-based learning with the bespoke goal of density estimation. Lastly, we establish a connection with energy-based models (EBMs) and point out some existing {\it implicit} applications of copulas in the field.

\paragraph{The DRE - MI - copula triangle:}
One of the many applications of DRE is to estimate MI, a measure of mutual dependence, computed as $KL(p||\prod_i p_i)$. This is identified as the negative copula entropy by \cite{davy2003copulas} and by \cite{ma2011mutual} (the canonical example of MI, {\it although not stated as such}, is the estimation of a Gaussian copula, {e.g.}~Section 6.2 in \cite{choi2022density}). The community has taken notice, in turn estimating MI using copulas: \cite{zeng2011estimation,ince2017statistical,lall2021stable}. The reverse direction is used by \cite{samo2021inductive} to estimate copulas with a constrained maximisation problem of MI. Thus, existing research has explored two edges of this triangle; DRE$\rightarrow$MI and MI$\leftrightarrow$copula. We connect the last edge, by using DRE to estimate copulas; DRE$\rightarrow$copula. In particular, our approach and \cite{choi2021featurized} share the use of an invertible mapping to facilitate the ratio estimation. 

\paragraph{Learning methods for copulas:} In \cite{mohamed2016learning}, three likelihood-free learning methods are discussed, with a shared goal of density estimation by comparison: [1] moment matching methods, [2] the maximum mean discrepancy (MMD), and [3] probabilistic classification (including Bregman and $f$-divergences). In the copula estimation community, likelihood-based training is prevalent. An exception is \cite{hofert2021quasi}, who use a generative moment matching network, a purely generative model, to learn a deep copula by matching the moments of samples against observations - exemplifying an application of [1] to copulas. Similarly, \cite{janke2021implicit} use the MMD as a learning metric for a generative network to replicate observed copula samples - showcasing the use of [2]. \cite{huk2023probabilistic} use the MMD as a learning metric for when MLE-based training is unavailable but still obtain a likelihood model. Relatedly, \cite{alquier2023estimation} estimate parametric copulas with the MMD, instead motivated by its robustness properties. Finally, the current paper is the first copula estimation method implementing [3]: the use of probabilistic classification. Our work is unique in constructing a deep copula model trained using a likelihood-free objective while still obtaining a density estimator - in contrast to the contributions of \cite{hofert2021quasi} and \cite{janke2021implicit}.

\paragraph{EBMs and other hidden copulas:}
EBMs \citep{NIPS2006_87f4d79e} are a class of unormalised models giving unconstrained values to data on $\mathcal{X}$, to maximise the values given to observations from $p$ and minimise the values anywhere else. They are often trained with contrastive objectives \citep{hinton2002training}, and in particular also employ probabilistic DRE \citep{gutmann2012noise,rhodes2020telescoping}. They can be interpreted as an extreme version of ratio copulas, aimed at capturing dependence in data, not only discriminating against $\prod_i p_i$ but the whole space of $\mathcal{X}$. Conversely, our ratio copula model is an EBM trained on the ratio of $p/\prod_i p_i$.
A notable use case of EBMs is modelling the latent space of variational autoencoders (VAE) \cite{kingma2013auto}, which is equipped with independent marginal densities. \cite{xiao2022adaptive} train an EBM with DRE to learn $p/\prod_i p_i$ for better performance, indirectly exemplifying the performance of ratio copulas! In fact, early classification DRE works advocate the use of $\prod_i p_i$ to learn $p$; \cite{hastie2009elements} p.479 and \cite{shi2006unsupervised} both unknowingly use a ratio copula model. Our paper thus consolidates their empirical work with a grounded derivation of this procedure, bridging the rich field of DRE with copula estimation.

\section{DISCUSSION}
\label{sec:diss}
In this work, we introduce the ratio copula as a novel framework for copula estimation. Our method fits a classifier between transformed copula samples and independent data on a Gaussian space, identifying dependence. We establish equivalences between popular copulas and classification tasks, providing new insights into their models. Empirically, our model is on par with the best copula methods for density evaluation as well as sampling. 

Future directions of our work can incorporate different losses \citep{menon2016linking} to fine-tune ratio copulas to focus on traits such as extreme dependence in the tails. An exciting direction is utilising other DRE methods for copula estimation, most notably linking copula density estimation with (time) score matching as in \cite{choi2022density}. Furthermore, some applications of copulas such as Bayesian Copula Networks \cite{elidan2010copula}, and conformal prediction \citep{messoudi2021copula,park2024semiparametric} desire to work with the copula distribution rather than the density. Extending our ratio copula formulation to obtain a model that parameterises CDFs is another meaningful direction.\par
Crucially, given the success copulas have enjoyed in related fields \citep{tagasovska2019copulas,czado2022vinefin,huk2024quasi} as methodological tools, further improvements would yield dividends across all of their applications.

\paragraph{Author Contributions:} David Huk originated the idea, designed the project, wrote the code, ran the experiments, wrote the paper, and derived the proofs. Ritabrata Dutta and Mark Steel jointly supervised the project, provided valuable feedback, and helped edit the paper.

\paragraph{Acknowledgements:} We thank all the reviewers for their helpful feedback. We also thank Alexis Derumigny, and Lorenzo Pacchiardi for helpful comments before the camera-ready version. David Huk is funded by the Center for Doctoral Training in
Mathematical Sciences at Warwick. Ritabrata Dutta is funded by EPSRC (grant nos. EP/V025899/1
and EP/T017112/1) and NERC (grant no. NE/T00973X/1).

\bibliography{main}

\section*{Checklist}



 \begin{enumerate}

 \item For all models and algorithms presented, check if you include:
 \begin{enumerate}
   \item A clear description of the mathematical setting, assumptions, algorithm, and/or model. [\textcolor{red}{Yes}] We include a notation part in Section \ref{sec:background}, and state assumptions at the start of relevant sections of the text.
   \item An analysis of the properties and complexity (time, space, sample size) of any algorithm. [\textcolor{red}{Yes}] We provide Lemma \ref{lemma:KL_gauss} to quantify the complexity of our task.
   \item (Optional) Anonymized source code, with specification of all dependencies, including external libraries. [\textcolor{red}{Yes}] Will be included in the additional material and published upon acceptance.
 \end{enumerate}

 \item For any theoretical claim, check if you include:
 \begin{enumerate}
   \item Statements of the full set of assumptions of all theoretical results. [\textcolor{red}{Yes}]
   \item Complete proofs of all theoretical results. [\textcolor{red}{Yes}] In the Appendix.
   \item Clear explanations of any assumptions. [\textcolor{red}{Yes}]     
 \end{enumerate}

 \item For all figures and tables that present empirical results, check if you include:
 \begin{enumerate}
   \item The code, data, and instructions needed to reproduce the main experimental results (either in the supplemental material or as a URL). [\textcolor{red}{Yes}]
   \item All the training details (e.g., data splits, hyperparameters, how they were chosen). [\textcolor{red}{Yes}]
         \item A clear definition of the specific measure or statistics and error bars (e.g., with respect to the random seed after running experiments multiple times). [\textcolor{red}{Yes}]
         \item A description of the computing infrastructure used. (e.g., type of GPUs, internal cluster, or cloud provider). [\textcolor{red}{Yes}]
 \end{enumerate}

 \item If you are using existing assets (e.g., code, data, models) or curating/releasing new assets, check if you include:
 \begin{enumerate}
   \item Citations of the creator If your work uses existing assets. [\textcolor{red}{Yes}]
   \item The license information of the assets, if applicable. [\textcolor{red}{Yes}]
   \item New assets either in the supplemental material or as a URL, if applicable. [\textcolor{red}{Yes}]
   \item Information about consent from data providers/curators. [\textcolor{red}{Not Applicable}]
   \item Discussion of sensible content if applicable, e.g., personally identifiable information or offensive content. [\textcolor{red}{Not Applicable}]
 \end{enumerate}

 \item If you used crowdsourcing or conducted research with human subjects, check if you include:
 \begin{enumerate}
   \item The full text of instructions given to participants and screenshots. [\textcolor{red}{Not Applicable}]
   \item Descriptions of potential participant risks, with links to Institutional Review Board (IRB) approvals if applicable. [\textcolor{red}{Not Applicable}]
   \item The estimated hourly wage paid to participants and the total amount spent on participant compensation. [\textcolor{red}{Not Applicable}]
 \end{enumerate}

 \end{enumerate}

\newpage
\appendix

%
%





%

%

\onecolumn
\aistatstitle{Supplementary Materials\\
Your copula is a classifier in disguise:\\
classification-based copula density estimation
}

\section{PROOFS}
\label{apdx:proofs}

\subsection{Theorem \ref{thm:cop}}
\label{apdx:thm}

We begin by stating assumptions needed for each part of the proof, followed by proving each of the three statements, namely [1] uniqueness, [2] consistency in the limit of infinite data from $p$, [3] asymptotic normality of the parameter estimate. This result is largely an application of \cite{gutmann2012noise} Theorems 1,2 and 3.

\paragraph{Assumption A1:}{ We begin by assuming that all marginals are known and, that they are continuous on the space $\mathcal{X}\in \mathbb{R}$ where $p(\mathbf{x})>0$ so that the copula exists by Sklar's theorem \citep{sklar}. We consider the loss $\mathcal{L}(\theta)$ from (\ref{eq:loss}), as in \cite{gutmann2012noise}. }

This is similar to the assumption taken for Sklar's theorem, where assuming continuity lets us derive the uniqueness of the copula. Positivity is required for the copula to exist in the first place.

 \paragraph{Assumption A2:} Our ratio copula model is well-specified, meaning $c(\mathbf{x})\in \mathcal{C}$.
 
 This is a classic assumption for deriving statistical results in model estimation, as done in the proof for the rate of convergence for simplified vine copula models \cite{nagler2016evading}. In our case, using MLP-based copula models gives us access to a much larger model class compared to simplified vines, making the discrepancy between the true model and the best model within our model class smaller, thereby reducing misspecification concerns.
\begin{proof}
As our marginals are positive whenever $p(\mathbf{x})>0$, the same goes for $q(\mathbf{x})=\prod_{i=1}^{d}p_i(x_i)$. This then applies to $\tilde p$ and $\tilde q$ too. Therefore, we satisfy the assumptions of Theorem 1 in \cite{gutmann2012noise}, which gives us the result namely that ${c}^{\text ratio}(\Phi^{-1}\circ\mathcal{P}(\mathbf{x});\boldsymbol{\theta}^*)$ is the only minimised of our loss. Further, as the model is well specified due to A2, we recover $c(\mathbf{x})=\tilde p(\mathbf{x})/\tilde q(\mathbf{x})= p(\mathbf{x})/ q(\mathbf{x})$.\end{proof}

 \paragraph{Assumption A3:} Keeping $\nu$ fixed, and increasing $T_p$ (meaning $T_q$ will also increase at a fixed constant rate), the supremum of the empirical approximation $\widehat{\mathcal{L}}(\boldsymbol{\theta})$ of Loss \ref{eq:loss} is such that, with the empirical loss defined as:
\begin{eqnarray}
\label{eq:loss_apdx}
\widehat{\mathcal{L}(\boldsymbol{\theta})} &=& \argmin_{\boldsymbol{\theta}\in\Theta}
    -\sum_{k=1}^{T_p} \ln \left(\frac{{c^{\text{ratio}}}(\Phi^{-1}\circ\mathcal{P}(\mathbf{x}^{k}_p);\boldsymbol{\theta})}{\nu+{c^{\text{ratio}}}(\Phi^{-1}\circ\mathcal{P}(\mathbf{x}^{k}_p);\boldsymbol{\theta})}\right)\\\nonumber
    && \quad \quad\,-\nu \cdot\sum_{k=1}^{T_q} \ln \left(\frac{\nu}{\nu+{c^{\text{ratio}}}(\Phi^{-1}\circ\mathcal{P}(\mathbf{x}^{k}_q) ; \boldsymbol{\theta})}\right),
\end{eqnarray}
 for $\mathbf{x}^{k}_{p} \sim \tilde p$ and $\mathbf{x}^{k}_q \sim \tilde q$, we have:
 \begin{equation}
     \text{sup}_{\boldsymbol{\theta}}|\widehat{\mathcal{L}(\boldsymbol{\theta})}-{\mathcal{L}(\boldsymbol{\theta})}|\rightarrow^{\mathbb{P}}0,\quad \text{when } T_p\rightarrow\infty. 
 \end{equation}

 \paragraph{Assumption A4:} 
Defining the matrix $\mathcal{I}_\nu$ as:
\begin{equation}
\mathcal{I}_\nu = \int \mathbf{g}(\mathbf{z}) \mathbf{g}(\mathbf{z})^T F_\nu(\mathbf{z}) \tilde p(\mathbf{z}) \, \mathrm{d} \mathbf{z},
\end{equation}

this matrix has full rank, for the quantities defined as below:

\begin{equation}
\mathbf{g}(\mathbf{z}) = \left. \nabla_{\boldsymbol{\theta}} \ln c^{\text{ratio}}(\mathbf{z}; \boldsymbol{\theta})\right|_{\boldsymbol{\theta}^{\star}}\cdot \tilde q(\mathbf{z}) , \quad \text{and} \quad
F_\nu(\mathbf{z}) = \frac{\nu \tilde q(\mathbf{z})}{\tilde p(\mathbf{z}) + \nu \cdot\tilde p(\mathbf{z})}.
\end{equation}
Assumptions A3 and A4 have their counterparts in maximum likelihood estimation, see {\it e.g.} \cite{wasserman2006all}. 

\begin{proof}
    As assumptions A1, A2, A3 and A4 imply that we satisfy assumption (a,b,c) of theorem 2 in \cite{gutmann2012noise}, we obtain its result, namely that $\widehat{\boldsymbol{\theta}}\rightarrow^{\mathbb{P}}\boldsymbol{\theta}^*$.
\end{proof}

Lastly, we obtain asymptotic normality of our parameter estimate.
\begin{proof}
    As we satisfy assumption A1-4, theorem 3 in \cite{gutmann2012noise} applies, and we obtain that $\sqrt{T_p}\left(\hat{\boldsymbol{\theta}}-\boldsymbol{\theta}^{\star}\right)$ is asymptotically normal with mean $\mathbf{0}$ and covariance matrix $\boldsymbol{\Sigma}$ wit expression:

\begin{equation}
\boldsymbol{\Sigma} = \boldsymbol{\mathcal{I}}_\nu^{-1} - \left(1 + \frac{1}{\nu}\right) \boldsymbol{\mathcal{I}}_\nu^{-1} \mathrm{E}\left(F_\nu \mathbf{g}\right) \mathrm{E}\left(F_\nu \mathbf{g}\right)^T \boldsymbol{\mathcal{I}}_\nu^{-1}
\end{equation}

where $\mathrm{E}\left(F_\nu \mathbf{g}\right) = \int F_\nu(\mathbf{z}) \mathbf{g}(\mathbf{z}) \tilde p(\mathbf{z}) \, \mathrm{d} \mathbf{z}$.

\end{proof}
In particular, as $\nu \rightarrow\infty$, $\Sigma$ becomes independent of the form of $\tilde q$, meaning that no matter the complexity of $c(\cdot)$ relative to a simple independence copula, with infinite noise samples, estimating it is only dependent on the shape of $p$ alone. Furthermore, under some mild assumptions detailed in \cite{geenens2017probit}, the resulting $\tilde p$ after applying transformation (\ref{eq:transform}) has unconstrained support and it, as well as its partial derivatives, will be bounded on $\mathbb{R}^d$. This is useful to ensure a well-behaved $\Sigma$ for asymptotic normality, since $\Sigma$ is defined in terms of the gradients of $\tilde p$.

\subsection{Lemma \ref{lemma:KL_gauss}}
\label{apdx:proof_KLgauss}
\begin{proof}
We first derive the KL divergence between the two Gaussians of the numerator and denominator in a Gaussian copula with covariance $\Sigma$.
    \begin{equation*}
        \begin{aligned}
&KL(\mathcal{N}_d(.;\mathbf{0},\Sigma)||\mathcal{N}_d(.;\mathbf{0},\mathbf{I}_d)) \\ &=\int p(\mathbf{x}) \cdot \left[\frac{1}{2} \log \frac{\left|\mathbf{I}_d\right|}{\left|\Sigma\right|}-\frac{1}{2}\left(\mathbf{x}-\mathbf{0}\right)^T \Sigma^{-1}\left(\mathbf{x}-\mathbf{0}\right)+\frac{1}{2}\left(\mathbf{x}-\mathbf{0}\right)^T \mathbf{I}_d^{-1}\left(\mathbf{x}-\mathbf{0}\right)\right] d \mathbf{x} \\
& =\frac{1}{2} \log \frac{1}{\left|\Sigma\right|}-\frac{1}{2} \mathbb{E}\left[\mathbf{x}^T\Sigma^{-1}\mathbf{x}\right] +\frac{1}{2} \mathbb{E}\left[\mathbf{x}^T \mathbf{x}\right] \\
& =\frac{1}{2} \log \frac{1}{\left|\Sigma\right|}-\frac{1}{2} \operatorname{tr}\left\{\Sigma^{-1} \Sigma\right\} +\frac{1}{2}\operatorname{tr}\left\{ \Sigma \right\} =\frac{1}{2}\log \frac{1}{\left|\Sigma\right|} 
\end{aligned}
    \end{equation*}
where the final equality follows from the fact that in a Gaussian copula, all dimensions marginally have unit variance, meaning $\Sigma$ has ones on the diagonal. Next, for $e_1,\ldots,e_d$ the eigenvalues of $\Sigma$ with $e_{min}$ the lowest of them, we get the desired bound
    \begin{equation*}
        \begin{aligned}
&KL(\mathcal{N}_d(.;\mathbf{0},\Sigma)||\mathcal{N}_d(.;\mathbf{0},\mathbf{I}_d)) = \frac{1}{2}\log \frac{1}{\left|\Sigma\right|}  \\ &= \frac{1}{2}\log \frac{1}{\prod_{i=1}^d e_i} \leq -\frac{d}{2}\log e_{min} .
\end{aligned}
    \end{equation*}
\end{proof}

\subsection{Proposition \ref{prop:cop_qda}}
\label{apdx:proofs_cop_qda}
\begin{proof}

A binary QDA classifier $h(\mathbf{z}:\mathbb{R}^d\mapsto\{0,1\})$ assigns class $y=1$ to $\mathbf{z}$ with a probability given by

\begin{equation}
    \mathbb{P}(y=1|\mathbf{z})=\frac{{\sqrt{(2 \pi)^d\left|\Sigma_1\right|}}^{-1} \exp \left(-\frac{1}{2}\left(\mathbf{z}-\boldsymbol{\mu}_1\right)^T \Sigma_1^{-1}\left(\mathbf{z}-\boldsymbol{\mu}_1\right)\right)}{{\sqrt{(2 \pi)^d\left|\Sigma_1\right|}}^{-1} \exp \left(-\frac{1}{2}\left(\mathbf{z}-\boldsymbol{\mu}_1\right)^T \Sigma_1^{-1}\left(\mathbf{z}-\boldsymbol{\mu}_1\right)\right)+{\sqrt{(2 \pi)^d\left|\Sigma_0\right|}}^{-1} \exp \left(-\frac{1}{2}\left(\mathbf{z}-\boldsymbol{\mu}_0\right)^T \Sigma_0^{-1}\left(\mathbf{z}-\boldsymbol{\mu}_0\right)\right)}
\end{equation}
and class $y=0$ otherwise, where $\mathbf{\mu_1},\mathbf{\mu_0}$ are the means and $\Sigma_1,\Sigma_0$ are the covariances of the Gaussian for classes 1 and 0 respectively. Next, we consider the following expression for this classifier
\begin{equation}
    \frac{\mathbb{P}(y=1|\mathbf{z})}{1-\mathbb{P}(y=1|\mathbf{z})}= \frac{{\sqrt{(2 \pi)^d\left|\Sigma_1\right|}}^{-1} \exp \left(-\frac{1}{2}\left(\mathbf{z}-\boldsymbol{\mu}_1\right)^T \Sigma_1^{-1}\left(\mathbf{z}-\boldsymbol{\mu}_1\right)\right)}{{\sqrt{(2 \pi)^d\left|\Sigma_0\right|}}^{-1} \exp \left(-\frac{1}{2}\left(\mathbf{z}-\boldsymbol{\mu}_0\right)^T \Sigma_0^{-1}\left(\mathbf{z}-\boldsymbol{\mu}_0\right)\right)}
\end{equation}
By assuming class 1 follows a centred Gaussian with covariance matrix $\Sigma$ and further assuming class 0 follows a standard Gaussian, that is setting $\mathbf{\mu_1}=\mathbf{0}, \mathbf{\mu_0}=\mathbf{0}$ and $\Sigma_1=\Sigma, \Sigma_0=\mathbf{I}_d$, we obtain

\begin{equation*}
    \frac{\mathbb{P}(y=1|\mathbf{z})}{1-\mathbb{P}(y=1|\mathbf{z})}=\frac{{\sqrt{(2 \pi)^d\left|\Sigma\right|}}^{-1} \exp \left(-\frac{1}{2}\mathbf{z}^T \Sigma_1^{-1}\mathbf{z}\right)}{{\sqrt{(2 \pi)^d}}^{-1} \exp \left(-\frac{1}{2}\mathbf{z}^T\mathbf{I}_d\mathbf{z}\right)}.
\end{equation*}
This can now be seen to be the ratio between the likelihood of a dependent Gaussian with mean $\mathbf{0}$ and covariance $\Sigma$ and a standard Gaussian, which is exactly equal to the expression of a Gaussian copula parameterised by $\Sigma$ when taking $\mathbf{z}=\Phi^{-1}\circ \mathcal{P}(\mathbf{x}) \in \mathbb{R}^d$ as inputs to both:
    \begin{equation*}
c_\Sigma(\Phi^{-1}\circ \mathcal{P}(\mathbf{x}))=\frac{\mathcal{N}_d(\Phi^{-1}\circ \mathcal{P}(\mathbf{x});\mathbf{0},\mathbf{\Sigma})}{\prod_{i=1}^d\mathcal{N}(\Phi^{-1}\circ P_i(x_i);0,1)}.
    \end{equation*}

\end{proof}

\subsection{Student's t copula}
\label{apdx:proof_student}
We provide here a result on non-Gaussian copulas, where we use the definition of a student's t-distribution with multiple degrees of freedom from \cite{ogasawara2024multivariate} (note that this is not the "usual" multivariate Student with one degree of freedom parameter, where the common latent mixing variable induces dependence even if correlations are zero):\\
\begin{proposition}{Consider a Student's t copula with density
$$
c_{\nu, P}^t(\mathbf{u})=\frac{f_{\nu, P}\left(F_\nu^{-1}\left(P_1(x_1)\right), \ldots, F_\nu^{-1}\left(P_d(x_d)\right)\right)}{\prod_{i=1}^d f_\nu\left(F_\nu^{-1}\left(P_i(x_i)\right)\right)},
$$
where for $\nu$ the degrees of freedom, $f_{\nu, P}$ is the d-dimensional student's t pdf with mean $\mathbf{0}$ and correlation matrix $P$, $F_\nu^{-1}$ is the inverse of a univariate standard student's t cdf, and $f_\nu$ is the density of the univariate standard $t$-distribution. The associated Bayes optimal classifier of this copula is found by the decision boundary between a student's t distribution and a student's t distribution with multiple degrees of freedom when all its degrees of freedom are equal to $\nu$, both densities with mean $\mathbf{0}$ but correlation matrices $P$ and $diag(P)$ respectively.}
\end{proposition}

\begin{proof}
The decision boundary of the above-mentioned classification problem will take the form of a ratio between the two densities. The numerator is simply $f_{\nu, P}$, matching the numerator of the copula density. The denominator of the copula is a product of independent marginal student-t densities with common degrees of freedom $\nu$. We note that a multivariate student's t does not factorise into $d$ marginals when its correlation matrix is diagonal. This can be understood through the equivalent representation of a student's t with $\mathbf{Z}\sim f_{\nu, P} \iff \mathbf{Z}=\mathbf{X}/\sqrt{Y/\nu}$ where $\mathbf{X}\sim\mathcal{N}(\mathbf{0},P)$ and $Y$ follows a Chi-squared distribution with $\nu$ degrees of freedom. As the same $Y$ is used across all dimensions, the resulting $\mathbf{Z}$ variables are not marginally independent. However, this is the case in a multivariate t-distribution with multiple degrees of freedom with a diagonal correlation matrix, as different Chi-squared variables are used for each dimension. When all Chi-squared variables share the same degree of freedom, the factorised form becomes that of the denominator of a student's t copula, concluding our proof.
\end{proof}

\subsection{Proposition \ref{prop:kde}}
\label{apdx:kde}
The proof consists of first identifying that a  KDE copula is a sum of ratios between Gaussians. Then, we use a similar reasoning to Proposition \ref{prop:cop_qda} for each summand to conclude the proof.
\begin{proof}
We begin with the expression of a Gaussian KDE copula $ \hat{c}_{T_p}(\Phi^{-1}\circ\mathcal{P}(\mathbf{x}))$ based on  $T_p$ kernel means at observations $\{(Z_1^k,\ldots,Z_d^k)\}_{k=1}^{T_p}$ for $Z_i^k=\Phi^{-1}\circ P_i(x_i^k),\, x_i^k\sim p_i$:
     \begin{equation}
        \hat{c}_{T_p}(\Phi^{-1}\circ\mathcal{P}(\mathbf{x}))=\frac{\Sigma_{k=1}^{T_p}\prod_{i=1}^{d}\{\mathcal{N}(\Phi^{-1}\circ P_i(x_i)-Z_i^{k}|0,b)\}}{T_p\cdot\mathcal{N}_d\left(\Phi^{-1}\circ \mathcal{P}(\mathbf{x})|\mathbf{0},\mathbf{I}_d\right)}.
    \end{equation}
    Notice that the expression in the numerator is a $d-$dimensional Gaussian density, as
    \begin{equation}
        \prod_{i=1}^{d}\{\mathcal{N}(\Phi^{-1}\circ P_i(x_i)-Z_i^{k}|0,b)\} = \mathcal{N}(\mathbf{z};\mathbf{Z}^k,b\cdot\mathbf{I}_d)
    \end{equation}
    where we write $\mathbf{z}=\Phi^{-1}\circ \mathcal{P}(\mathbf{x})$ and $\mathbf{Z}^k=(Z_1^k,\ldots,Z^k_d)$. It then ensues that 
    \begin{equation}
        \hat{c}_{T_p}(\Phi^{-1}\circ \mathcal{P}(\mathbf{x}))= \frac{\sum_{k=1}^{T_p}}{T_p}\frac{\mathcal{N}(\mathbf{z};\mathbf{Z}^k,b\cdot\mathbf{I}_d)}{\mathcal{N}_d\left(\mathbf{z}|\mathbf{0},\mathbf{I}_d\right)}
    \end{equation}
    showing a sum of ratios of Gaussians with means $\mathbf{Z}^k$ and $\mathbf{0}$ and variances $b\cdot\mathbf{I}_d$ and $\mathbf{I}_d$ for the numerator and denominator respectively. \\
    Following the same reasoning as in Proposition \ref{prop:cop_qda}, each such term coincides with the ratio of posterior class probabilities of a binary QDA classifier for two Gaussian classes with the aforementioned means and variances, having labels $y=1$ for the numerator and $y=0$ for the denominator class. Concretely, denoting each such QDA classifier as $h^k(\mathbf{z};\mu_1^k,\mu_0^k,\Sigma_1,\Sigma_0)$, shortening the notation to $h^k(\mathbf{z})$, with means $\mu_1^k=\mathbf{Z}^k,\mu_0^k=\mathbf{0}$ but shared variances $\Sigma_1=b\cdot\mathbf{I}_d,\Sigma_0=\mathbf{I}_d$, we obtain the following ratio:

    \begin{equation}
         \delta^k(\mathbf{z}):=\frac{h^k(\mathbf{z})}{1-h^k(\mathbf{z})}= \frac{{\sqrt{(2 \pi)^d\left|b\cdot\mathbf{I}_d\right|}}^{-1} \exp \left(-\frac{1}{2}\left(\mathbf{z}-\mathbf{Z}^k\right)^T b\cdot\mathbf{I}_d^{-1}\left(\mathbf{z}-\mathbf{Z}^k\right)\right)}{{\sqrt{(2 \pi)^d\left|\mathbf{I}_d\right|}}^{-1} \exp \left(-\frac{1}{2}\left(\mathbf{z}-\mathbf{0}\right)^T \mathbf{I}_d^{-1}\left(\mathbf{z}-\mathbf{0}\right)\right)}.
    \end{equation}
    Now, considering an average of such QDA decision boundaries, denoted as $\delta_{T_p}(\mathbf{z}):=\frac{\sum_{k=1}^{T_p}}{T_p}\delta^k(\mathbf{z})$, we get:
\begin{equation}
    \delta_{T_p}(\mathbf{z}) = \frac{\sum_{k=1}^{T_p}}{T_p}\frac{{\sqrt{(2 \pi)^d\left|b\cdot\mathbf{I}_d\right|}}^{-1} \exp \left(-\frac{1}{2}\left(\mathbf{z}-\mathbf{Z}^k\right)^T b\cdot\mathbf{I}_d^{-1}\left(\mathbf{z}-\mathbf{Z}^k\right)\right)}{{\sqrt{(2 \pi)^d\left|\mathbf{I}_d\right|}}^{-1} \exp \left(-\frac{1}{2}\left(\mathbf{z}-\mathbf{0}\right)^T \mathbf{I}_d^{-1}\left(\mathbf{z}-\mathbf{0}\right)\right)} = \hat{c}_{T_p}(\Phi^{-1}\circ \mathcal{P}(\mathbf{x})).
\end{equation}
Thus, $\delta_{T_p}(\mathbf{z})$ and $\hat{c}_{T_p}(\Phi^{-1}\circ \mathcal{P}(\mathbf{x}))$ coincide, showing that a KDE copula with Gaussian kernels is equivalent to a mixture of QDA decision boundaries with mixture specific means per class, but same variances per class across mixtures, concluding our proof.
\end{proof}

\subsection{Corollary \ref{coro:vine}}
\label{apdx:proof_vine}

\begin{proof}
We begin by stating the simplified vine copula density for a $d$-dimensional vector $(u_1, \ldots, u_d) \in [0,1]^d$ and $i,j \in \{1,\ldots,d\}$. For more details on vine copulas, the reader is referred to Appendix \ref{apdx:copulas}.
\begin{equation}
    \mathbf{c}(u_1, \ldots, u_d) = \prod_{i\neq j}^{d(d-1)/2} c_{ij}(P_{i|\mathcal{S}_{ij}}(x_{i|\mathcal{S}_{ij}}), P_{j|\mathcal{S}_{ij}}(x_{j|\mathcal{S}_{ij}})),
\end{equation}
The conditioning sets $\mathcal{S}_{ij}$ are determined by the specific vine structure, and are a possibly empty subset of dimensions excluding $i,j$. Each of the $d(d-1)/2$ copulas in the decomposition takes two arguments $P_{i|\mathcal{S}_{ij}}(x_{i|\mathcal{S}_{ij}}), P_{j|\mathcal{S}_{ij}}(x_{j|\mathcal{S}_{ij}})$ as inputs, both marginally uniform in $[0,1]$, and outputs their bivariate copula density. Based on the modelling assumption chosen for those bivariate densities $c_{i,j}$, we follow our previous results to deduce separate equivalences for the final vine model.

\paragraph{Part 1: Simplified vine with bivariate Gaussian copulas.}
When using bivariate copula building blocks, the vine equation becomes: 

\begin{equation}
    \mathbf{c}(u_1, \ldots, u_d) = \prod_{i\neq j}^{d(d-1)/2} \frac{\mathcal{N}_2\left(\Phi^{-1}\left(P_{i|\mathcal{S}_{ij}}(x_{i|\mathcal{S}_{ij}})\right),\Phi^{-1}\left(P_{j|\mathcal{S}_{ij}}(x_{j|\mathcal{S}_{ij}})\right);\mathbf{0},\mathbf{\Sigma}\right)}{\mathcal{N}(\Phi^{-1}\left(P_{i|\mathcal{S}_{ij}}(x_{i|\mathcal{S}_{ij}})\right);{0},{1})\cdot\mathcal{N}(\Phi^{-1}\left(P_{j|\mathcal{S}_{ij}}(x_{j|\mathcal{S}_{ij}})\right);{0},{1})}.
\end{equation}
As $P_{i|\mathcal{S}_{ij}}(x_{i|\mathcal{S}_{ij}}), P_{j|\mathcal{S}_{ij}}(x_{j|\mathcal{S}_{ij}})$ are marginally uniform, by the probability integral transformation, this expression is a ratio of two bivariate Gaussians with means $\mathbf{0}$ and covariances $\mathbf{\Sigma},\mathbf{I}_2$ for the numerator and denominator respectively. From Proposition \ref{prop:cop_qda}, it follows that each term in the vine decomposition is a QDA decision boundary, giving the result.

\paragraph{Part 2: Simplified vines with Gaussian KDE bivariate copulas.}
Similarly, each bivariate KDE copula is a ratio between a mixture of bivariate kernels and a bivariate standard Gaussian, with some inputs on the $\mathbb{R}^2$ Gaussian latent space that are marginally Gaussian due to the probability integral transformation. Due to the form of a Gaussian KDE copula, Proposition \ref{prop:kde} applies to each of the $d(d-1)/2$ bivariate copulas involved in the vine decomposition (up to replacing the observations $\{(Z_1^k,\ldots,Z_d^k)\}_{k=1}^{T_p}$ by new ${Z_1^\prime}^k:=\Phi^{-1}\circ P_{i|\mathcal{S}_{ij}}(x_{i|\mathcal{S}_{ij}})$ which include the conditioning implied by vines), meaning that the simplified vine density is then composed as a product of $d(d-1)/2$ bivariate classifiers, each being an average of $T_p$ decision boundaries obtained by QDA on a 2-dimensional Gaussian problem.
\end{proof}

\subsection{Equivalences for parametric copulas in terms of classifiers}
\label{apdx:cop_class}
We can use our framework to obtain the classifiers implied by known copulas. Indeed, 
$$c(\mathbf{P(\mathbf{x})})=\frac{h(\mathbf{x})}{1-h(\mathbf{x})} \iff h(\mathbf{x})=\frac{c(\mathbf{P(\mathbf{x})})}{1+c(\mathbf{P(\mathbf{x})})}$$
which we use to empirically illustrate the classifiers implied by various copula families by using their densities.\par

\begin{figure}
    \centering
    \includegraphics[width=0.8\linewidth]{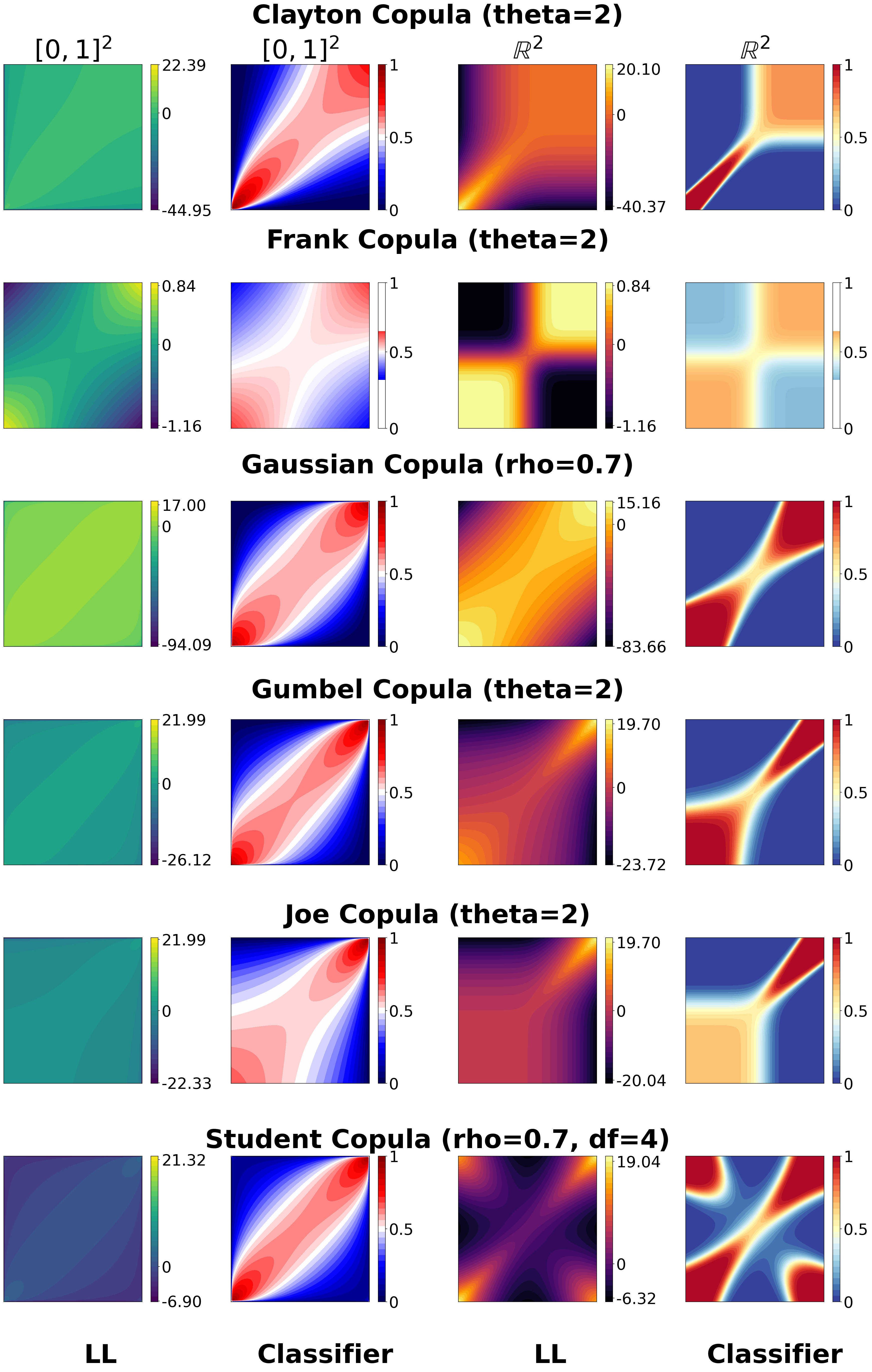}
    \caption{Example of six parametric bivariate copulas and their equivalent classifiers $h^*(\mathbf{x})$. The first two columns show the copula log-likelihoods (LL) and their implied classifiers on the copula $[0,1]^2$ scale, while the last two columns show the same quantities but on the Gaussian $\mathbb{R}^2$ scale.}
    \label{fig:cop_class}
\end{figure}

Figure \ref{fig:cop_class} shows six different bivariate copulas and their classifiers on the copula $[0,1]^2$ space as well as the Gaussian space via a dimension-wise inverse Gaussian transform. A notable benefit of representing copulas by their classifier counterparts is an improved clarity of the shape; the LL values $\log c(U,V)$ in the plots of all but the Frank copula explode at the corners which makes them hard to evaluate, whereas the classifier versions $h(u,v)$ depict clear shapes in accordance with those of copulas. This can be seen on the Gaussian space plots, where the classifier identifies the diagonal as being high-likelihood regions for the dependent data and places increasingly high probability on the diagonal as the distance from the centre (and so from the independent factorised density) increases. The classifier's increasingly high probability is expected, as for instance, the Gaussian copula density is unbounded in the corners on the diagonal. Using our classifier interpretation also exemplifies the theoretical results from Section \ref{sec:equivalences}. Further, the similarity of the classifier to the shape of the copula is expected as the copula LL can be understood as a difference of the LL of the numerator and denominator of a copula i.e. $\log c(P_1(c_1),P_2(x_2))=\log p(x_1,x_1)-\log ( p_1(x_1)\cdot p_2(x_2))$, which is what the classifier is trained to distinguish.

\section{COPULA MODELS}
\label{apdx:copulas}

Copulas are widely used in statistics and machine learning for modelling joint probability distributions. They enable the construction of joint densities through a two-step estimation process. First, the marginal densities are estimated as if they were independent. Then, the copula is modelled to capture the dependence structure between the variables. This approach is grounded in Sklar's theorem:
\begin{theorem}[\cite{sklar}]
\label{thm:sklar}
    Let $\mathbf{P}$ be a $d$-dimensional distribution function with continuous marginal distributions $P_1, P_2, \ldots, P_d$. Then there exists a copula distribution $\mathbf{C}$ such that for all $\mathbf{x} = (x_1, x_2, \ldots, x_d) \in \mathbb{R}^d$:
\begin{equation}
\mathbf{P}(x_{1}, \ldots, x_{d}) = \mathbf{C}(P_{1}(x_{1}), \ldots, P_{n}(x_{d})).
\end{equation}
And if a probability density function (pdf) is available:
\begin{equation}
\mathbf{p}(x_{1}, \ldots, x_{d}) = p_{1}(x_{1}) \cdot \ldots \cdot p_{d}(x_{d}) \cdot \mathbf{c}(P_{1}(x_{1}), \ldots, P_{d}(x_{d})),
\end{equation}
where $p_{1}(x_{1}), \ldots, p_{d}(x_{d})$ are the marginal pdfs, and $\mathbf{c}(P_{1}(x_{1}), \ldots, P_{d}(x_{d}))$ is the copula pdf.
\end{theorem}

If the marginal distributions are continuous, the copula is unique. This allows the estimation of the joint distribution \( p(\mathbf{x}) \) to be broken down into two steps: first, estimating the marginal distributions \(\{p_i\}_{i=1}^d\), and then modeling the copula \(c(u_1,\ldots ,u_d)\), where \(u_i := P_i(x_i)\) for \(i \in \{1,\ldots,d\}\), are the images of the \(x_i\) under the CDF of each dimension. By applying CDF transformations to the marginals, the copula becomes agnostic to differences between dimensions, such as axis scaling, and focuses solely on capturing the dependence structure among them.

\subsection{Gaussian copula}
\label{apdx:copulas_gauss}
A widely used parametric copula model is the \emph{Gaussian copula}, which assumes that the dependence between dimensions follows that of a multivariate Gaussian distribution with mean \(\mathbf{0}\) and covariance matrix \(\Sigma\). It transforms uniform marginals into Gaussian variables and then models their dependence using the Gaussian distribution. Therefore, the diagonal of $\Sigma$ only has ones by construction as the uniform variables of the copula are transformed with an inverse standard Gaussian CDF. The copula density is given by:
\[
c(u_1,\ldots, u_d) = \frac{\mathcal{N}_d(\Phi^{-1}(u_1),\ldots,\Phi^{-1}(u_d);\mathbf{0},\Sigma)}{\prod_{i=1}^d \mathcal{N}(\Phi^{-1}(u_i);0,1)}.
\]
Here, \(\Phi^{-1}\) is the inverse cumulative distribution function (CDF) of the standard normal distribution, and \(\mathcal{N}_d\) denotes the multivariate normal density. The parameters of the Gaussian copula are the off-diagonal entries of the covariance matrix \(\Sigma\), which capture the dependence between dimensions. For \(d=2\), there is a single parameter to estimate, representing the correlation between the two variables.

\subsection{Vine copulas}
\label{apdx:vine}
A \emph{vine copula} is a flexible method for modelling dependencies in high-dimensional data by decomposing a $d$-dimensional copula into a sequence of $d(d-1)/2$ bivariate copulas through structured conditioning \citep{bedford2001probability}. This allows for efficient estimation by modelling pairs of variables instead of the full joint distribution \citep{nagler2016evading}. The decomposition is based on the following identity, which follows from Sklar's theorem:
\begin{equation}
\label{eq:cop_identity}
    p_{a|b}(x_a|x_b) = c_{a,b}(P_a(x_a),P_b(x_b)) \cdot p_a(x_a)
\end{equation}
where $a$ and $b$ are subsets of the dimensions $\{1, \ldots, d\}$. Vine copulas rely on a conditional factorization $\mathbf{p}(x_1, \ldots, x_d) = \prod_{i=1}^d p_{i|<i}(x_i|x_{<i})$, repeatedly applying~\eqref{eq:cop_identity} to split the joint density into $d$ marginal densities and $\frac{d(d-1)}{2}$ bivariate pair copulas.

A full vine copula model for the joint copula density can be written as:
\begin{equation}
\label{eq:qbvine}
    \mathbf{c}(u_1, \ldots, u_d) = \prod_{i\neq j}^{d(d-1)/2} c_{ij}(P_{i|\mathcal{S}_{ij}}(x_{i|\mathcal{S}_{ij}}), P_{j|\mathcal{S}_{ij}}(x_{j|\mathcal{S}_{ij}})|\mathcal{S}_{ij}),
\end{equation}
where each pair copula depends on the conditional distributions $P_{i|\mathcal{S}_{ij}}$ and $P_{j|\mathcal{S}_{ij}}$, and $\mathcal{S}_{ij}$ is a subset of dimensions excluding $i$ and $j$. The set $\mathcal{S}$ conditions the copulas to capture higher-order dependencies.

In a simplified vine copula, the conditioning on $\mathcal{S}$ is ignored, leading to a model that is easier to work with:
\begin{equation}
\label{eq:simplified_vine}
    \mathbf{c}(u_1, \ldots, u_d) = \prod_{i\neq j}^{d(d-1)/2} c_{ij}(P_{i|\mathcal{S}_{ij}}(x_{i|\mathcal{S}_{ij}}), P_{j|\mathcal{S}_{ij}}(x_{j|\mathcal{S}_{ij}})),
\end{equation}
where the pair copulas are now modelled as unconditional bivariate copulas, simplifying the dependency structure.

For instance, in the case of a 3-dimensional random vector $(U, V, W)$, the joint copula density can be factorized as:
\begin{equation}
c_{U,V,W}(u,v,w)\,=\,c_{U,V}(u,v)\cdot c_{V,W}(v,w)\cdot c_{U,W|V}\bigg(C_{U|V}(u|v),\,C_{W|V}(w|v)\bigg|v\bigg)\, ,
\end{equation}
where $c_{U,V}(u,v)$ and $c_{V,W}(v,w)$ are the bivariate copulas for pairs $(U,V)$ and $(V,W)$, and $c_{U,W|V}$ captures the conditional dependence between $U$ and $W$ given $V$. The conditional copula depends on the conditional CDFs $C_{U|V}$ and $C_{W|V}$, making the model more complex but allowing greater flexibility in capturing dependencies. With a simplified vine, which assumes that the pairwise copulas do not change with conditioning, the factorization becomes:
\begin{equation}
c_{U,V,W}(u,v,w) = c_{U,V}(u,v) \cdot c_{V,W}(v,w) \cdot c_{U,W|V}(C_{U|V}(u|v), C_{W|V}(w|v)).
\end{equation}

\section{PRACTICAL DETAILS}
\label{apdx:exp}


Here, we provide more details on running experiments and using the ratio copula in practice. We show our algorithm for fitting the ratio copula model in Algorithm \ref{alg:ratio_copula}, where orange parts are to be included optionally to produce a guided Gaussian ratio copula model instead of a simple ratio copula.

\begin{algorithm}[h]
\caption{Training a Ratio Copula }
\label{alg:ratio_copula}
\begin{tcolorbox}[colback=orange!10!white, colframe=orange!50!black, title=]
Optionally, with \textcolor{orange}{Gaussian-Guided} ratio copula.
\end{tcolorbox}
\textbf{Require:} Dataset $\mathcal{D}_p \sim p(\mathbf{x})$ and parameterised ratio copula ${c}^\text{ratio}(\mathbf{z};\boldsymbol{\theta}):\mathbb{R}^d \mapsto\mathbb{R}$;
\begin{enumerate}
\item Dataset $\mathcal{D}_{\tilde p} = \{\Phi^{-1}\circ \mathcal{P}(\mathcal{D}_p)\}$ on Gaussian scale;
\item \begin{tcolorbox}[colback=orange!10!white, colframe=orange!50!black, width=0.9\linewidth, sharp corners] Estimate sample covariance  $\Sigma\leftarrow \text{Cov}(\mathcal{D}_{\tilde p})$; \end{tcolorbox}
\textbf{While} $\boldsymbol{\theta}$ not converged \textbf{do}:
    \begin{itemize}
    \item Sample $\mathbf{z}_{\tilde p} \in \mathcal{D}_{\tilde p}$ and $\mathbf{z}_q \sim \mathcal{N}_d(.;\mathbf{0},\mathbf{I}_d)$; 
    \item \begin{tcolorbox}[colback=orange!10!white, colframe=orange!50!black, sharp corners] Instead, sample $\mathbf{z}_{\tilde p} \in \mathcal{D}_{\tilde p}$ and $\mathbf{z}_{\tilde q} \sim \mathcal{N}_d(.;\mathbf{0},\Sigma)$; \end{tcolorbox}
    \item Compute ${c}^\text{ratio}(\mathbf{z}_{\tilde p};\boldsymbol{\theta})$ and ${c}^\text{ratio}(\mathbf{z}_{\tilde q};\boldsymbol{\theta})$;
    \item Calculate loss $\mathcal{L}(\boldsymbol{\theta})$ using Equation \eqref{eq:loss};
    \item Update parameters $\boldsymbol{\theta} \leftarrow \boldsymbol{\theta} - \nabla_{\boldsymbol{\theta}} \mathcal{L}(\boldsymbol{\theta})$;
    \end{itemize}
\end{enumerate}
\textbf{Return:} Ratio copula model ${c}^\text{ratio}(\mathbf{z};\boldsymbol{\hat{\theta}})$.\\
\begin{tcolorbox}[colback=orange!10!white, colframe=orange!50!black, width=\linewidth, sharp corners] \textbf{Return:} Gaussian-Guided ratio copula model $c^\text{GG}(\mathbf{z};\boldsymbol{\hat{\theta}}):={c}^\text{ratio}(\mathbf{z};\boldsymbol{\hat{\theta}})\cdot \frac{\mathcal{N}_d(\mathbf{z}|\mathbf{0},\mathbf{\Sigma})}
{\mathcal{N}_d(\mathbf{z}|\mathbf{0},\mathbf{I}_d)}$. \end{tcolorbox}
\end{algorithm}

\paragraph{Preprocessing:} For each experiment, we start with data on $\mathcal{X}$ and need to apply an estimate of the CDFs to obtain data on a copula $[0,1]^d$ scale. We use the empirical CDF as a consistent approximator of the true CDF for all experiments. While this leads to an approximate version of samples on $[0,1]^d$, we treat these samples as samples from the true copula density, using the same empirical CDF per dataset across all methods fitted to make results comparable. Since only the copula LL matters for showing a copula model is performant, using the same CDF across methods ensures that the comparison is fair and focuses on the modelling capabilities of each model, not on the CDF approximation.

\paragraph{Hyperparameters and network architectures:}
In our experiments, for the simplified vine copula, we use a TLL constant copula as the bivariate building block, corresponding to a bivariate KDE copula. The TLL constant copula requires a bandwidth $b$ for the kernel. We followed standard literature on KDE estimation by choosing Silverman's rule of thumb for the form of the bandwidth \citep{silverman2018density}, based on the sample size $n$ in the training data, and the dimension of the data $d$:
\begin{equation}
    b=\left(\frac{n  (d + 2)}{4}\right)^{(-1 / (d + 4))}.
\end{equation}
However, in the Digits experiment, this choice resulted in numeric overflow in the LL evaluation (likely due to the low number of sampled compared to the dimension), leading us to choose the bandwidth $b$ with a grid search using cross-validation on training data. Further, simplified vine copulas can be {\it truncated} at a given level $t\leq d$, meaning that the number of bivariate copulas is limited at the first $t(t-1)/2$ such copulas in the decomposition. This is equivalent to setting all bivariate copulas past a certain level of conditioning as independent copulas ({\it i.e} equal to $1$). This is a common approach to reduce the computational burden of fitting a vine copula \citep{czado2022vine}. We use a full set of $d(d-1)/2$ bivariate copulas for the Digits experiment, but use a truncation level of $50$ for the MNIST experiment, in order to keep computational times of the vine copula similar to other models. 

For the IGC, we follow the experimental setup of \cite{janke2021implicit} in their original paper. We use a fully connected MLP, with an initial layer taking as input Gaussian noise, with dimension $3 d$, followed by two fully connected layers with width $100$, with a final connection from the second fully connected layer to the $d$-dimensional output. This architecture is used for all experiments with the exception of MNIST where a third fully connected hidden layer was used. We show the parameter number per experiment in Table \ref{tab:arch_IGC}.

\begin{table}[htbp]
\centering
\begin{tabular}{|l|c|}
\hline
\textbf{Experiment Name} & \textbf{Number of Parameters} \\
\hline
2D Image & 11,002 \\
Digits & 35,864 \\
MNIST & 324,584 \\
\hline
\end{tabular}
\caption{Number of Parameters in the IGC copula for Different Experiments.}
\label{tab:arch_IGC}
\end{table}

Finally, our ratio copula model follows a similarly simple architecture to the IGC, to show its general applicability. We specifically chose architectures with a similar parameter count to make comparisons fair. In the Digits and MNIST examples, we use convolutional layers following a standard classifier architecture for image data. We provide a table describing our ratio copula architecture in Tables \ref{tab:arch_ratio_2dcop}, \ref{tab:arch_ratio_2dimg}, \ref{tab:arch_ratio_digits}, \ref{tab:arch_large_ratio_digits}, \ref{tab:arch_ratio_mnist}, and \ref{tab:arch_large_ratio_mnist}. We use a different number of independent noise samples in the optimisation procedure of Loss (\ref{eq:loss}) based on the complexity of the task and the desired trade-off between computational time and statistical efficiency. In the 2D parametric copula example, we use the whole 5000 samples and 50,000 noise samples (but these are fixed across epochs). In the 2D image experiment, we use 10,000 true samples and 100,000 noise samples per epoch. In the Digits experiment, we use the whole 898 training samples and use 89,800 noise samples per batch. In the MNIST example, we use the whole data of size 30,000 and 300,000 independent samples.

\begin{table}[htbp]
\centering
\begin{tabular}{|l|c|c|}
\hline
\textbf{Layer (type)} & \textbf{Output Shape} & \textbf{Param \#} \\
\hline
Linear               & [-1, 1, 100]       & 300      \\
ReLU + Residual      & [-1, 1, 100]       & 0        \\
Linear               & [-1, 1, 100]       & 10,100   \\
ReLU + Residual      & [-1, 1, 100]       & 0        \\
Linear               & [-1, 1, 100]       & 10,100   \\
ReLU + Residual      & [-1, 1, 1]         & 0        \\
Linear               & [-1, 1, 1]         & 101      \\
\hline
\textbf{Total params:} & \multicolumn{2}{c|}{\textbf{20,601}} \\
\hline
\end{tabular}
\caption{Summary of the ratio copula architecture used in the 2D parametric copula examples. Here, a simple ratio copula model is used.}
\label{tab:arch_ratio_2dcop}
\end{table}

\begin{table}[htbp]
\centering
\begin{tabular}{|l|c|c|}
\hline
\textbf{Layer (type)} & \textbf{Output Shape} & \textbf{Param \#} \\
\hline
Linear     & [-1, 1, 100]       & 300      \\
\textit{ReLU} + Residual & [-1, 1, 100]       & 0        \\
Linear     & [-1, 1, 100]       & 10,100   \\
\textit{ReLU} + Residual & [-1, 1, 100]       & 0        \\
Linear     & [-1, 1, 100]       & 10,100   \\
\textit{ReLU} + Residual & [-1, 1, 100]       & 0        \\
Linear     & [-1, 1, 100]       & 10,100   \\
\textit{ReLU} + Residual & [-1, 1, 100]       & 0        \\
Linear     & [-1, 1, 100]       & 10,100   \\
\textit{ReLU} + Residual & [-1, 1, 100]       & 0        \\
Linear     & [-1, 1, 100]       & 10,100   \\
\textit{ReLU} + Residual & [-1, 1, 100]       & 0        \\
Linear     & [-1, 1, 1]         & 101      \\
\hline
\textbf{Total params:} & \multicolumn{2}{c|}{\textbf{50,901}} \\
\hline
\end{tabular}
\caption{Summary of the ratio copula architecture used for the 2D image data. Here, a simple ratio copula model is used.}
\label{tab:arch_ratio_2dimg}
\end{table}

\begin{table}[htbp]
\centering
\begin{tabular}{|l|c|c|}
\hline
\textbf{Layer (type)} & \textbf{Output Shape} & \textbf{Param \#} \\
\hline
Conv2d     & [-1, 64, 4, 4]     & 640      \\
LeakyReLU  & [-1, 64, 4, 4]     & 0        \\
Conv2d     & [-1, 64, 2, 2]     & 36,928   \\
LeakyReLU  & [-1, 64, 2, 2]     & 0        \\
Flatten    & [-1, 256]          & 0        \\
Linear     & [-1, 1]            & 257      \\
\hline
\textbf{Total params:} & \multicolumn{2}{c|}{\textbf{37,825}} \\
\hline
\end{tabular}
\caption{Summary of the ratio copula architecture used in Digits experiments. We use a GG+ratio model for Digits.}
\label{tab:arch_ratio_digits}
\end{table}

\begin{table}[htbp]
    \centering
    \begin{tabular}{|l|c|c|}
        \hline
        \textbf{Layer (type)} & \textbf{Output Shape} & \textbf{Param \#} \\
        \hline
        Conv2d          & [-1, 64, 4, 4]     & 640     \\
        LeakyReLU      & [-1, 64, 4, 4]     & 0       \\
        Conv2d         & [-1, 64, 2, 2]     & 36,928  \\
        LeakyReLU      & [-1, 64, 2, 2]     & 0       \\
        Flatten        & [-1, 256]          & 0       \\
        Linear         & [-1, 256]          & 65,792  \\
        LeakyReLU      & [-1, 256]          & 0       \\
        Linear         & [-1, 1]            & 257     \\
        \hline
        \textbf{Total params:} & \multicolumn{2}{c|}{\textbf{103,617}} \\
        \hline
    \end{tabular}
    \caption{Summary of the GG+Ratio \textit{large} copula architecture used in Digits experiments.}
    \label{tab:arch_large_ratio_digits}
\end{table}

\begin{table}[htbp]
\centering
\begin{tabular}{|l|c|c|}
\hline
\textbf{Layer (type)} & \textbf{Output Shape} & \textbf{Param \#} \\
\hline
Conv2d    & [-1, 64, 14, 14]   & 640      \\
LeakyReLU  & [-1, 64, 14, 14]   & 0        \\
Conv2d   & [-1, 64, 7, 7]     & 36,928   \\
LeakyReLU  & [-1, 64, 7, 7]     & 0        \\
Flatten    & [-1, 3136]         & 0        \\
Linear     & [-1, 64]           & 200,768  \\
LeakyReLU  & [-1, 64]           & 0        \\
Linear    & [-1, 32]           & 2,080    \\
LeakyReLU  & [-1, 32]           & 0        \\
Linear    & [-1, 1]            & 33       \\
\hline
\textbf{Total params:} & \multicolumn{2}{c|}{\textbf{240,449}} \\
\hline
\end{tabular}
\caption{Summary of the ratio copula architecture used in MNIST experiments. We use a GG+ratio model for MNIST.}
\label{tab:arch_ratio_mnist}
\end{table}

\begin{table}[htbp]
    \centering
    \begin{tabular}{|l|c|c|}
        \hline
        \textbf{Layer (type)} & \textbf{Output Shape} & \textbf{Param \#} \\
        \hline
        Conv2d          & [-1, 64, 14, 14]   & 640     \\
        LeakyReLU      & [-1, 64, 14, 14]   & 0       \\
        Conv2d         & [-1, 64, 7, 7]     & 36,928  \\
        LeakyReLU      & [-1, 64, 7, 7]     & 0       \\
        Flatten        & [-1, 3136]         & 0       \\
        Linear         & [-1, 128]          & 401,536 \\
        LeakyReLU      & [-1, 128]          & 0       \\
        Linear         & [-1, 64]           & 8,256   \\
        LeakyReLU      & [-1, 64]           & 0       \\
        Linear         & [-1, 64]           & 4,160   \\
        Linear         & [-1, 32]           & 2,080   \\
        Linear         & [-1, 1]            & 33      \\
        \hline
        \textbf{Total params:} & \multicolumn{2}{c|}{\textbf{453,633}} \\
        \hline
    \end{tabular}
    \caption{Summary of the GG+ratio \textit{large} model architecture used in MNIST experiments.}
    \label{tab:arch_large_ratio_mnist}
\end{table}

\paragraph{Optimisation:}
For the IGC and ratio copula models, we employ the Adam optimiser \citep{diederik2014adam} in Pytorch with the base settings, to demonstrate the ease of use of our model, as in \cite{janke2021implicit}. For the 2D parametric copula experiments, we optimise these models for 500 steps, for the 2D image experiments we use 100,000 optimisation steps, for the Digits dataset we use 500 steps and we use 1000 steps for MNIST to reflect the complexity of the task. For the ratio copula, at each epoch, we resample a new set of independent noise (either from $\mathcal{N}(\cdot;\mathbf{0},\mathbf{I}_d)$ or from a correlated Gaussian $\mathcal{N}(\cdot;\mathbf{0},\Sigma)$ in the Gaussian guided ratio copula). Therefore, for the ratio copula, following the analysis in Theorem \ref{thm:cop} and \cite{gutmann2012noise}, there is a trade-off available for our method between computation time and statistical performance; as we optimise the model for more epochs and sample more independent noise, the better our estimate of the true ratio copula. Therefore, while we limit the number of epochs to the same number as the IGC, our method could benefit from longer training times.

\paragraph{Diagnostics for overfitting} Other than using Lemma \ref{lemma:KL_gauss}, an easily available diagnostic is to look at the loss during training. Since the copula naturally overlaps independent with dependent samples, we should always observe a positive loss. If it is too close to 0, then the classifier is overfitting and the two densities are too far apart \citep{rhodes2020telescoping}. Furthermore, using a normalising constant estimator also provides a diagnostic of fit: copula densities are already normalised by definition, hence if correctly estimated, the normalising constant will be 1. If any of these symptoms arise, it signals we should employ a better DRE method for estimating the ratio copula. For instance, switching from a ratio copula to a Gaussian-Guided ratio copula.

\paragraph{Sampling from the ratio copula:} 
As our ratio copula model has an analytical density, we employ Markov chain Monte Carlo methods to obtain samples. Specifically, we use the Hamiltonian Monte Carlo scheme \citep{neal2011mcmc}, from the implementation in the pyhmc python package \citep{pyhmc}. For each sample, we initiate the chain at a random sample from a standard independent Gaussian on the latent Gaussian space, and implement 100 gradient-informed MCMC steps to ensure the chain has reached the main mass of the density. We then select the next step after that as the sample from our model. As in {\it e.g.} KDE or Gaussian copula models, this sampling is done from $p$ on the latent Gaussian space, meaning that we multiply our final ratio copula by $\mathcal{N}(\cdot;\mathbf{0},\mathbf{I}_d)$, and target the resulting density with our MCMC sampling scheme. The samples obtained by the MCMC scheme are consequently on a Gaussian scale, meaning we apply a Gaussian CDF to map them to the copula $[0,1]^d$ space.

\paragraph{Metrics:}
We use the LL, as measured by the output of our ratio model's log-density on held-out test data. Similarly, we use the python implementation of \cite{vatter20220} of simplified vine copula models as well as bivariate copulas to output LL values. For the Wasserstein 2 metric, we make use of the implementation of \cite{py_ot}. For all experiments, we measure the W2 using 500 samples from models, except for MNIST where we use 25 samples per model.

\paragraph{Computation:} 
All experiments were run on a single Intel Xeon Platinum 8260 (Cascade Lake) CPU. We show a table of runtimes for a single run of each model on experiments in Table \ref{tab:runtime}. In the 2D image experiment of Figure \ref{fig:2d_img} of the main text (2D IMG), vine denotes a TLL quadratic copula, and in the 2D parametric copulas (2D copulas) the vine denotes TLL copula runtimes (constant, linear and quadratic TLL take about the same time in those settings). For the 2D copulas, the first time for the ratio copula is a simple ratio copula with an MLP model (2:30 mins), while the second time (20s) is for fitting a logistic classifier as the ratio copula model. With 2 runs for 2D IMG, $25\cdot3$ runs for 2D copulas, and 10 runs for Digits and MNIST, our total computation time for the experiments is 65 hours on a single CPU.

\begin{table}[htbp]
\centering
\begin{tabular}{@{}lcccc@{}}
\toprule
\textbf{Model\textbackslash Data}   & \textbf{2D IMG} & \textbf{2D copulas} & \textbf{Digits} & \textbf{MNIST} \\ \midrule
Ratio            & 4h 30 mins  & 2:30 mins/20 s & 10 mins        & 105 mins       \\
IGC              & 3h          & -            & 2 mins         & 10 mins        \\
Vine             & 5 mins $^*$      &  15 s $^*$         & 4 mins         & 140 mins       \\
 \bottomrule
\end{tabular}
\caption{Time taken for fitting and sampling for a single run of different models on datasets from Section \ref{sec:Experiments}. The asterisks (*) signify that the TLL bivariate copula from the vine package was used instead of a multivariate vine.}
\label{tab:runtime}
\end{table}

\subsection{Additional experiments on losses for tail modelling}
 In \cite{menon2016linking}, authors propose losses that emphasize high density-ratio values. In copulas, under our ratio copula perspective, this corresponds to extreme tail dependence.\par

\begin{wrapfigure}{r}{0.45\textwidth}  
    \centering
    \includegraphics[width=\linewidth]{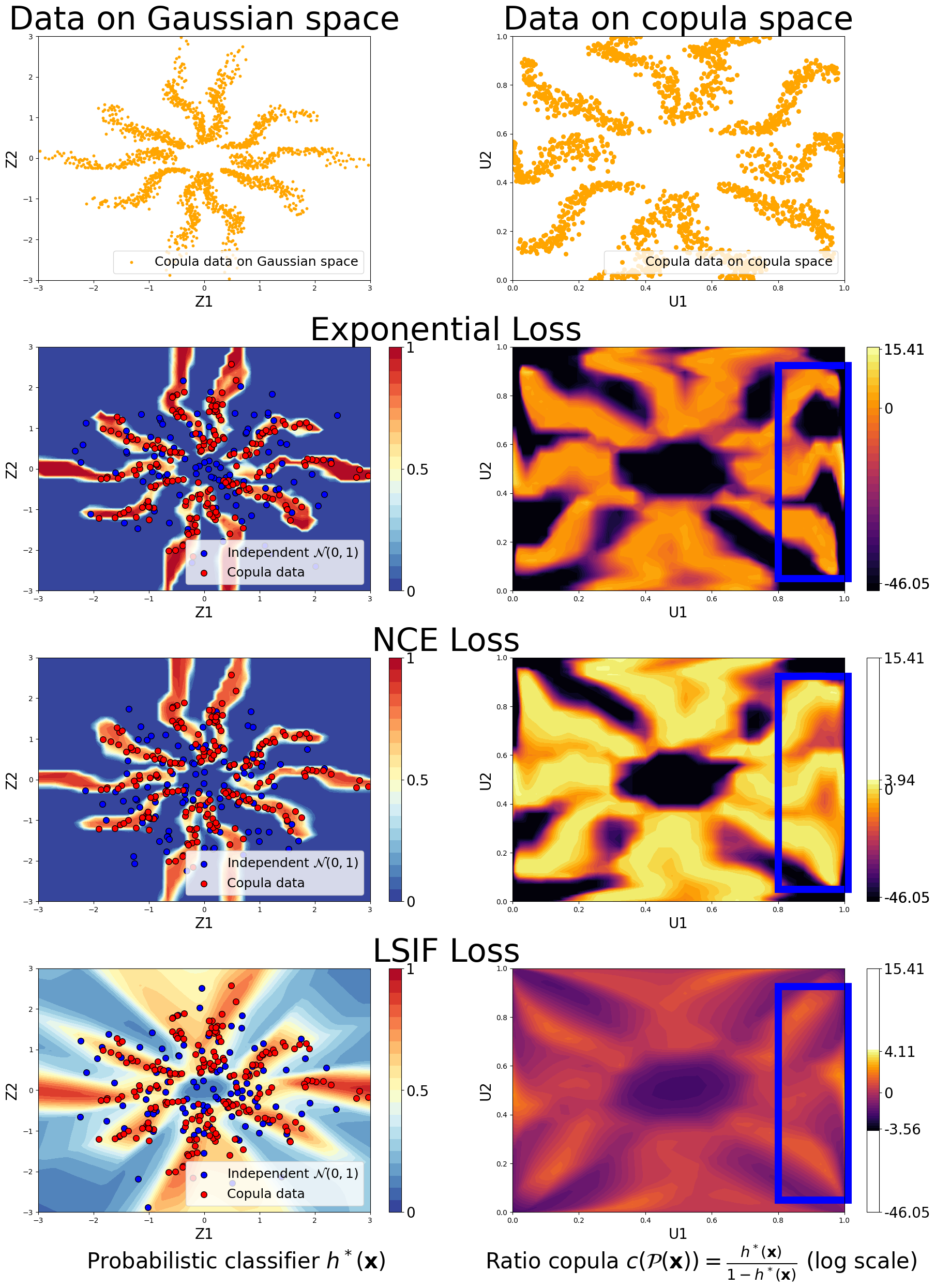}
    \caption{Same Ratio copula model trained with losses that emphasise tail behaviours. The exponential loss obtains a better separation of likelihood regions in the tails (boxed in blue) while the NCE loss conflates tail events and the LSIF loss is too diffuse.}
    \label{fig:losses}
\end{wrapfigure}

We exemplify the effect of different losses which place more or less emphasis on the tails for estimating ratio copulas. The Noise Contrastive Estimation (NCE) loss of \cite{gutmann2012noise} we use emphasizes training according to a weight in terms of the copula value $c(\mathcal{P}(\mathbf{x}))$ as $w(c)=1/(c^2 + c)$, while the exponential loss has $w(c)=1/c^{(3/2)}$ and the least
squares importance fitting (LSIF) loss \citep{JMLR:v10:kanamori09a} has $w(c)=1$, both latter losses placing a greater focus on tails. We illustrate this behaviour on a two-dimensional example shown in Figure \ref{fig:losses}. For observed data on the first row, we fit the same ratio copula model using different loss functions, showing the resulting classifiers and log-likelihoods on the left and right columns respectively. The exponential loss better captures tail events (boxed in blue), with sharper densities and better separation of observation "rays", while NCE conflates them. The higher LL values given to data-regions highlight the advantage of the exponential loss. The LSIF loss is harder to optimize which makes its density more diffuse. We hereby demonstrate another advantage of connecting copulas to DRE, allowing practitioners to make use of already developed tools, in this instance to improve tail modelling.

\subsection{Experiment on bivariate copulas from pictures in Figure \ref{fig:2d_img}}
\label{apdx:exp_2dimg}
The data to fit was obtained by normalising the pixel values of a monochromatic picture, to sum up to 1 across the image. We then sample with replacement following the probabilities given by the brightness of each pixel and add minute Gaussian noise to dequantize samples. This constitutes our dataset on $\mathbb{R}^2$, for which we estimate marginals (with empirical CDFs) and use them to finally obtain copula samples (the plots from the first column, with white dots for each sample) on $[0,1]^2$ used to fit copula models. Notice how the letters of the bottom row have different shades of white: this is because of the transformation to the copula space. Our model recovers these nuances, see the horizontal bars of each "T" which have a larger density than the vertical bars. In Figure \ref{fig:2d_img}, we display on $[0,1]^2$, going from left to right: the observed dataset, the LL estimated by the ratio copula, samples from the IGC, and the LL estimated from the TLL.

\subsection{High dimensional experiments}
\label{apdx:exp_digits_mnist}
\begin{figure}[h]
    \centering
    \includegraphics[width=\linewidth]{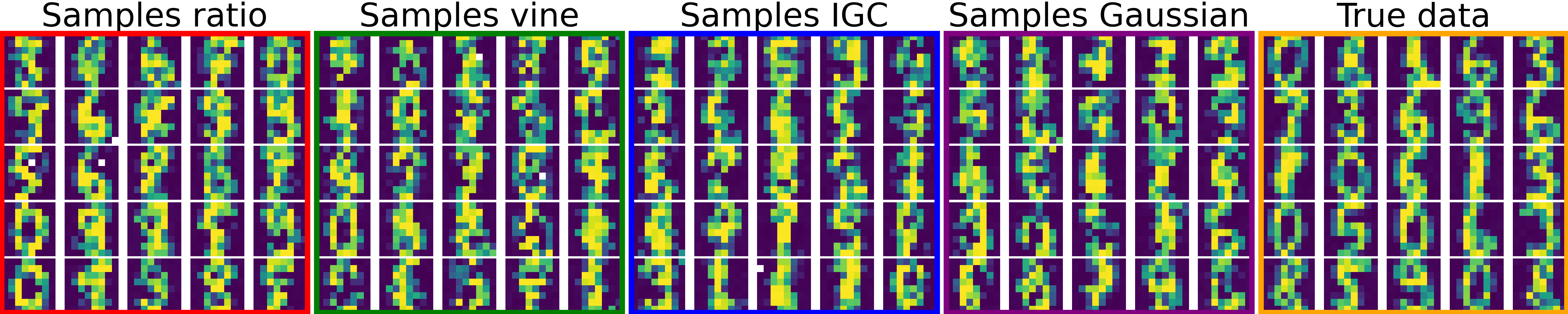}
    \caption{Random samples from all models from the Digits experiment. Samples are obtained by random sampling from the copula models and transformed back to the data space. The copula samples shown are not related to the true data, meaning the digits shown are not meant to match between models and true data.}
    \label{fig:digits_samples}
\end{figure}

\begin{figure}[h]
    \centering
    \includegraphics[width=\linewidth]{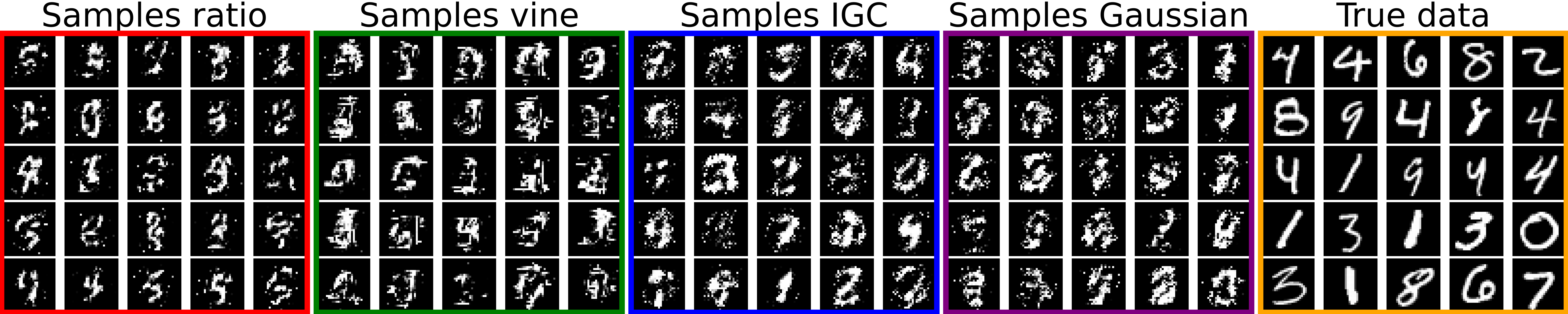}
    \caption{Random samples from all models from the MNIST experiment. Samples are obtained randomly from the copula models and transformed back to the data space. The Ratio copula obtains visually compelling samples while other copula models are more dispersed.}
    \label{fig:mnist_samples}
\end{figure}

The two high-dimensional datasets used are the Digits dataset \citep{alimoglu1997combining} and MNIST \citep{lecun1998mnist}. They both show handwritten digits in $8\cdot 8$ or $28\cdot 28$ format. As both these datasets represent complex dependencies, as measured using Lemma \ref{lemma:KL_gauss} from the main text, we use a GG+ratio copula model. We show results for a simpler (GG+Ratio) and larger version (GG+Ratio \textit{large} see Tables \ref{tab:arch_large_ratio_digits} and \ref{tab:arch_large_ratio_mnist}), to showcase the potential for scalability of ratio copulas. In Figure \ref{fig:digits_samples}, we show random samples from the simpler ratio copula, the simplified vine copula, the IGC and the Gaussian copula as well as true test data. We show samples from the same copula classes fitted on the MNIST data in Figure \ref{fig:mnist_samples}. Samples shown are obtained from the copula model and then transformed back to the data space using the empirical CDF as described in the preprocessing paragraph. We note that the samples shown are random across tiles, meaning that we are not showing how well a model imitates a given target digit. We are exploring the learned distribution implied by the model after training on the data samples, doing so by showing a collection of uncurated random samples from any given model.

For the Digits dataset, all models perform well. Arguably, the Gaussian copula sometimes fails to correctly capture the dependence to the same level as what is present in the true data. Other models perform similarly, with differences being hard to notice. The vine copula does have some samples where the dependence on the image follows certain {\it lines} along pixels, likely due to the bivariate decomposition used in vines causing pixels nearby to be correlated by ignoring longer dependencies in space due to the simplifying assumption. In particular, the ratio copula produces samples with high fidelity to the true data, consistently across all depicted samples.

For MNIST, the IGC and Gaussian copula samples are more blurry and imprecise, while the vine copula has samples with distinct {\it lines} forming the picture. The reason for the blurriness of the Gaussian copula is believed to be its less precise dependence structure, being strictly diagonal for the Gaussian. The reason for the blurriness in the IGC is not clear, but we believe the non-convexity of the MMD as a loss might make it harder for the model to recover the optimal dependence structure, instead settling at a local optimum. The reason for the apparent lines in the vine is likely associated with the recursive decomposition of a simplified vine copula, where bivariate copulas propagate dependence in a chain-like fashion, from neighbour to neighbour. Given that exact model selection is extremely hard for simplified vines, this dependence structure will likely be only approximately correct, causing some parts of the numbers to be misrepresented. This might be further amplified by the simplifying assumption, causing the model to not necessarily capture all the dependence present in the data. Again, the ratio copula consistently produces high quality samples, on par with the best of competing methods.

\vfill


\end{document}